
\documentstyle{amsppt}
\catcode`\@=11
\def\logo@{}
\catcode`\@=13
\TagsOnRight
\parindent=8 mm
\magnification 1200
\hsize = 6.25 true in
\vsize = 8.7 true in
\hoffset = .2 true in
\parskip=\medskipamount
\baselineskip=16pt
%
\def\rom{\roman}

\def \smaller {\eightpoint}
\def \wt {\widetilde}
\def \wh {\widehat}

\def \ra {\rightarrow}

\def \lra {\longrightarrow}
\def \lmt {\longmapsto}
\def \wc {\overset {\wedge}\to{,}}
\def \a {\alpha}
\def \d {\delta}

\def \s {\sigma}

\def \th {\theta}

\def \o {\omega}

\def\bl{{\;\scriptscriptstyle{{}^\rfloor}\;}}
\def \smskip{\smallskip}
%

\def\gr#1{{\goth #1}}



	\def\grg{{\gr g}}


\def\eps{\varepsilon}

\def\nchi{\hbox{\raise 2.5pt\hbox{$\chi$}}}





\def\hbar{\bar h}



\topmatter
\title
Symplectic Geometries on $T^*\wt{G}$, Hamiltonian Group Actions
 and Integrable Systems${}^\dag$
\endtitle
\footnote""{${}^{\dag}$ Research supported in part by the
       Natural Sciences and Engineering Research Council of Canada and
       the National Science Foundation.}
\rightheadtext{Hamiltonian Group Actions on $T^*\wt{G}$
 and Integrable Systems}
\leftheadtext{J. Harnad and B.A. Kupershmidt}
\author
 J. Harnad${}^1$ and B.A. Kupershmidt${}^2$
\endauthor
\endtopmatter
\footnote""{${}^1$Department of Mathematics and
  Statistics, Concordia University, Montr\'eal, P.Q. and \newline
  Centre de Recherches Math\'ematiques,
  Universit\'e de Montr\'eal, C.P. 6128-A,
  Montr\'eal, P.Q. H3C 3J7}
\footnote""{${}^2$ The University
of Tennessee Space Institute, Tullahoma, TN  37388, U.S.A.}
\bigskip \bigskip
\centerline{\bf Abstract}
\bigskip
\baselineskip=10pt
\centerline{
\vbox{
\hsize= 5.5 truein
{\smaller
Various Hamiltonian actions of loop groups $\wt G$ and  of the algebra
$\text{diff}_1$ of first order differential operators in one variable
are defined on the cotangent bundle $T^*\wt G$.
The moment maps generating the $\text{diff}_1$ actions are
shown to factorize through those generating the loop group actions,
thereby defining commuting diagrams of Poisson maps to the duals of the
corresponding centrally extended algebras. The maps are then used to derive a
number of infinite  commuting families of Hamiltonian flows that are
nonabelian generalizations of the dispersive water wave hierarchies.
 As a further application, sets of pairs of generators of the
 nonabelian mKdV hierarchies are shown to give a commuting hierarchy on
$T^*\wt G$ that contain the WZW system as its first element.}}}
\document
\baselineskip=16pt
\bigskip \bigskip
%
\bigskip
\line{\indent {\bf Introduction}\hfill}
\medskip
Integrable $1+1$-dimensional systems have long been recognized as closely
connected with loop algebras (see e.g. {\bf [AHP, DS, RS, FNR]}).
The r\^ole played by loop {\it groups} in the Hamiltonian setting is in a
sense more fundamental, but also more subtle
{\bf [DJMK, SW, RS, W1, W2, H, HK1]}. For many integrable systems the
underlying phase space may be taken as the cotangent bundle $T^*LG$ of a loop
group $LG$. However, the symplectic structure is not necessarily the
canonical one; it may more generally be a member of a  $1$-parameter
family obtained by shifting the canonical form by a
multiple of the $2-$cocycle on the corresponding loop algebra
$L\grg$ associated to an $Ad$--invariant metric on $\grg$.
It is this shift that
leads to a phase space structure involving the centrally extended loop
algebra.  Reduction of $T^*LG$ under left or right translations leads to
the space $LG/G = \{\text{Maps } S^1 \ra G\}/\{\text{constant loops}\}$
(a coadjoint orbit in the dual $L\grg^{\wedge *}$ of the centrally extended
loop algebra), with the natural symplectic form related to the $2$-cocycle
(cf. {\bf [PS, H]}).
By combining such reductions with the moment maps generating the natural
action of $\text{diff } S^1$ on $LG$ and $L\grg$, one obtains Poisson
maps relating integrable hierarchies of the KdV and nonabelian mKdV
type {\bf [Ku2, HK1, HK2]}.  In this context, the moment map provides a
nonabelian  generalization of the well known Miura map.

In the present work  we show,  using the  moment maps generating certain
 infinitesimal  Hamiltonian actions of the algebra $\text{diff}_1$ of
first order differential  operators in one variable on the nonreduced phase
 space $T^*LG$, how some new, nonstandard
Lax equations determining  infinite commuting families of flows are derived.
These systems are related to the  integrable hierarchy  of dispersive
water wave  (DWW) equations {\bf[Ku1]} in a way analogous to the relation
between the  nonabelian modified KdV and KdV hierarchies. They will be
referred to here as the $\grg$-mDWW ($\grg$-modified DWW) systems.
The relevant phase space for the latter is the dual
$(L\grg \oplus L\grg)^{\wedge *}$
of the centrally extended direct sum of two copies of the loop algebra
$L\grg$. (Alternatively, nonperiodic, rapidly decreasing boundary
conditions may be allowed, in which case the corresponding groups and
algebras will denoted $\wt{G}$ and $\wt{\grg}$, respectively.) A second
version, involving a different Hamiltonian action of $\text{diff}_1$ on
$T^*LG$, leads to systems on the phase space  $(L\grg \dotplus L\grg_A)^{*}$
(or $(\wt{\grg} \dotplus \wt{\grg})^{\wedge *}$), where the subscript $A$
denotes abelianization, and the sum is semi-direct. These will be referred
to  as the $\grg$-m$^2$DWW (second $\grg$-modified DWW) systems.

The relevant  family of symplectic structures on $T^*LG$ (or  $T^*\wt G$)
is defined in Section 1a. The rest of Section 1 is devoted to a systematic
study of  the various Hamiltonian actions of $\wt G$ on
$T^*\wt G$, $(\wt{\grg} \oplus \wt{\grg})^{\wedge *}$ and
$(\wt{\grg} \dotplus \wt{\grg}_A)^{\wedge *}$, as well as the  that of
the group $\Cal D_1 \ltimes \Cal D_0$ corresponding to $\text{diff}_1$ .
The associated moment maps are derived and shown to form commuting triplets
of Poisson maps  into  $(\wt{\grg} \oplus \wt{\grg})^{\wedge *}$  (or
$(\wt{\grg} \dotplus \wt{\grg}_A)^{\wedge *}$) and
 $\text{diff}_1^{\ \wedge*}$,
where $\text{diff}_1^{\ \wedge}$  is a member of a $3$-parameter family of
central extensions of the algebra $\text{diff}_1$. Using these Poisson
maps in a way analogous to the generalized Miura map, various integrable
systems associated with the dispersive water wave hierarchy are constructed
in Section 2. Since the KdV and nonabelian mKdV hierarchies  may be recovered
by restricting every second flow to a certain invariant
manifold, the DWW and  $\grg$-mDWW systems may be viewed as generalizations of
the KdV and  nonabelian mKdV systems, which  are more usually defined on
the phase spaces  $(\text{diff } S^1)^{\wedge *}$ and
$\wt{\grg}^{\wedge *}$ (or $L\grg^{\wedge *}$), respectively.
The Poisson maps constructed in Section 1 are recast in Section 2 in the
language of differential  algebras, Hamiltonian matrices and Hamiltonian
maps. In addition to the DWW, $\grg$-mDWW and $\grg$-m$^2$DWW hierarchies,
 we also obtain integrable hierarchies on  $T^*LG$ via pullbacks under the
 appropriate Hamiltonian maps. As a further application of these results,
we show that there exists a natural notion of higher WZW systems associated to
pairs of KdV systems that are related to the separated  left and right
translational modes of the WZW system.
\bigskip
\line{{\bf 1. \quad Hamiltonian Group Actions on $T^*\wt G$}\hfil}
\medskip
\line{1a. \quad{\it Symplectic Structures on $T^*\wt G$}\hfil}
\medskip
Let $G$ be a Lie group with Lie algebra $\grg$ on which an Ad-invariant
 scalar product $\Cal B:\grg \times \grg \ra \Bbb R$ is defined. We denote by
$\wt G$ the group of smooth maps $g: \Bbb R \ra G$, with pointwise
multiplication, and by $\wt \grg$ its
Lie algebra, consisting similarly of maps $X: \Bbb R \ra \grg$.  If the
periodicity conditions $g( \s + 2\pi) =  g(\s), \ X(\s + 2\pi) = X(\s)$ are
added, $\wt G$ and $\wt \grg$ become the loop group $LG$ and loop
algebra $L \grg$, respectively.  For the nonperiodic case we also require $g
\in \wt G, \ X \in \wt \grg$ to satisfy $L^2$ boundary conditions such that
the integral
$$
\align
\langle X, Y \rangle & := \int  \Cal{B}(X(\s), Y(\s))d\s  \tag {1.1}
\\
X , Y & \in \wt \grg
\endalign
$$
converge over $\Bbb R$, defining an Ad-invariant scalar product on $\wt
\grg$, and the same integral converge when the pair $(X,Y)$  is replaced by
$(g^{\prime}g^{-1}, h^{\prime}h^{-1})$ for any pair of group elements
$g, h \in \wt{G}$. For $L \grg$ the integral in (1.1) is understood as
evaluated over a
period.
\medskip
The scalar product $\Cal B$ gives an identification between $\grg$ and its dual
space ${\grg}^{*}$, while $<\  ,\  >$ gives an identification of $\wt
\grg$ as a dense subspace of $\wt {\grg}^{*}$.  For our purposes, when
speaking of $\wt {\grg}^{*}$, we shall really only mean elements in this
dense subspace, and hence the same notation
$$
\align
\langle U, X \rangle := & \int \Cal{B}(U(\s), X(\s))d\s  \tag {1.2}
\\
U \in \wt {\grg}^{*} \ ,& \quad   X \in \wt \grg
\endalign
$$
will be used to denote the dual pairing $\wt \grg^ * \times \wt \grg \ra
\Bbb R$.  The 2-cocycle
$$
\align
c: \wt \grg \times \wt \grg & \ra \Bbb R
\\
c(X, Y) & :=\langle {X , Y^{\prime}} \rangle \ ,  \tag {1.3} \\
Y^{\prime}&:=\frac{dY}{d\s}\ ,
\endalign
$$
is used to define the centrally extended algebra, denoted \  $\wt
{\grg}^{\wedge}$ (or $L \grg^{\wedge}$), identified as the space $\wt
\grg + \Bbb R$ (respectively $L \grg + \Bbb R$), with Lie brackets:
$$
\align
\lbrack (X, a) , (Y, b) \rbrack  & := \ ( \lbrack X, Y \rbrack , \ \langle
{X , Y'} \rangle )  \tag {1.4}
\\
X , Y \in \wt \grg , & \quad a, b \in \Bbb R \ .
\endalign
$$
The dual space $\wt {\grg}^{\wedge  *}$ is again identified with
$\wt \grg^* + \Bbb R \sim \wt \grg + \Bbb R$, with dual pairing
$$
\align
\langle (U, a) , (X , b) \rangle  & := \langle U , X \rangle + ab
\tag
{1.5}
\\
 (U,a) \in \wt \grg^{\wedge  *} &, \quad (X , b) \in \wt \grg^{\wedge} \ .
\endalign
$$
Denoting by $ad_X ^*$ and $Ad_g^*$ the coadjoint representations of
elements $X \in \wt \grg  , \  g \in \wt G$ in the algebra and group,
respectively, the extended coadjoint representation on $\wt
\grg^{\wedge *}$ is given by the formulae:
$$
\align
\wh{ad}^*_{(X, a)}  (U, b) & = ({\text{ad}}^*_X U + bX', 0)
 \tag {1.6a}
\\
\wh{Ad}^*_{\hat g}(U, b) & = ({\text{Ad}}^*_g U + bg'g^{-1}, b) \ .
\tag{1.6b}
\endalign
$$
Although the centrally extended group $\wt G^\wedge \ra \wt G$ with Lie
algebra $\wt \grg ^\wedge$ is a nontrivial line or circle bundle over $\wt
G$ (cf\.  {\bf[PS]}), its action on $\wt \grg ^{\wedge *}$ depends only on
the projection of elements $\hat g \in \wt G^{\wedge}$ to their image
$g \in {\wt G}$, as indicated in (1.6b).

{}From (1.4), (1.5), the Lie Poisson bracket on $\wt \grg ^{\wedge *}$ is
determined by the formula
$$
\{ \langle \  .\  \vert (X , a) \rangle ,  \langle \ .\   \vert (Y , b) \rangle
\}
\mid_{(U , e)}
= \langle U , \lbrack X , Y \rbrack \rangle +  e c (X , Y)   \tag {1.7}
$$
where $\langle \ .\  \vert (X , a) \rangle$ denotes the linear functional on
 $\wt \grg^{\wedge *}$ with value  $\langle (U,e) , (X , a) \rangle$ at
$(U,e)\in  \wt \grg^{\wedge *}$,
corresponding to pairing with the element $(X , a)  \in \wt \grg^{\wedge}$.

Now consider the cotangent bundle $T^{*} \wt G$, identified through left
translations with the product $\wt G \times \wt \grg^{*}$.  Denoting a
typical element by the pair $(g \in \wt G , \mu \in \wt \grg^{*})$, the
canonical symplectic form on  $T^{*} \wt G$ may be expressed as
$$
\o_0 = - \delta \langle \mu , g^{-1} {\delta g} \rangle  \ ,   \tag {1.8}
$$
where $\delta$ denotes the functional differential and $g^{-1} {\delta g}$
is the left invariant Maurer-Cartan form on $\wt G$.  In the following, a
more general family of symplectic forms will be needed (cf\.  {\bf[H]}),
defined by the formula:
$$
\o_k :=- \delta \langle \mu , g^{-1} {\delta g} \rangle + \frac {k}{2}
\langle \delta (g^{-1} g^{\prime}) \wc  g^{-1} {\delta g} \rangle  \ ,
\tag {1.9}
$$
where $k$ is a real parameter.  Here, $g^{-1} g^{\prime}$ is viewed as the
element of $\wt \grg$  whose value at $\s \in S^1$ is obtained by left
translation of the tangent vector  $g^{\prime} (\sigma) \in T_{g(\sigma)} G$ to
the identity $e \in G$.   It is shown in {\bf [H]}  that $\o_{k}$ is
closed and weakly nondegenerate. The corresponding Poisson brackets are
defined by the  formulae (cf. {\bf FT]}):
$$
\align
\lbrace f_{1} , f_{2} \rbrace & = 0 \qquad \qquad \tag  {1.10a}
\\
\{f, \langle \ .\ \vert X \rangle \} \mid_{(g , \mu)} &=
\int \delta f (gX) d\s
 \qquad \tag  {1.10b}
\\
\{ \langle \ .\ \vert X \rangle , \langle \ .\  \vert Y \rangle \}
\mid_{(g ,\mu)}
& = - \langle \mu + k g^{-1} g^{\prime} , [ X , Y ] \rangle + kc(X , Y) \ ,
 \tag  {1.10c}\\
X, Y  &\in \wt{\grg}
\endalign
$$
where $f , f_{1} , f_{2}$ are functionals on $\wt G \times \wt \grg^{*}$
depending only on the first factor $g \in \wt G$,
$\langle \ .\ \vert X\rangle$ is
the linear form on the second factor corresponding to the algebra element
$X\in \wt{\grg}$ and the RHS of (1.10b)
signifies evaluation of the functional differential $\delta f$ on the vector
$gX \in T_{g} \wt G$ obtained by left translation of $X \in \wt \grg \sim
T_{e} \wt G$ to $g \in \wt G$ (i.e. evaluation of the corresponding left
invariant vector field on $f$). The Poisson brackets (1.10a-c) follow from
the following formula for the Hamiltonian vector field  $X_H$ of an arbitrary
smooth functional $H$ defined over $T^*\wt{G}$:
$$
X_H= \langle g \frac{\d H}{\d\mu},\frac{\d}{\d g}\rangle
+\langle \left[\mu+ k g^{-1}g', \frac{\d H}{\d \mu} \right]
+k\left( \frac{\d H}{\d \mu}\right)'- \frac{\d H}{\d g} g,
\frac{\d {}}{\d \mu} \rangle\
 \tag{1.10d}
$$
(so that  $X_H \bl \o_k = \d (H)= \langle \frac{\d H}{\d g}, \d g\rangle
+\left< \frac{\d H}{\d \mu}, \d  \mu \right>$).

\bigskip

\line{1b. \quad{\it Hamiltonian $\wt G \times \wt G$ Action on $T^*\wt
G$}\hfil}
\medskip
The canonical lift of the {\it right} translation action of $\wt G$ to $T^* \wt
G$ is
given by:
$$
R_{h^{-1}} : (g , \mu) \lmt (g h^{-1} , h \mu h^{-1})\ , \tag {1.11}
$$
where the inverse $h^{-1}$ is used to make this a left action.  This action
leaves the symplectic form $\o_k$ invariant for all $k$.  To lift the
{\it left} translation action so as to preserve $\o_k$, it is necessary
to modify the usual canonical lift by defining:
$$
L_h : (g , \mu) \lmt (hg , \mu + kg^{-1}h^{-1}h^{\prime}g) \ .  \tag {1.12}
$$
Substitution in eq\. (1.9) shows that this does leave $\o_k$ invariant and
composing these maps shows that (1.12) defines a left action, commuting
with the right translation action (1.11).  The following result summarizes
the Hamiltonian properties of these actions.
\proclaim
{Proposition 1.1} The $R$ and $L$ actions
(1.11), (1.12) are generated in terms of Hamiltonian flows by the
moment maps:
$$
\align
J^L : T^* \wt G & \lra \wt \grg^*
\\
J^L : (g , \mu) & \lmt {g \mu g^{-1}}  \tag {1.13a}
\\
J_k^R : T^* \wt G & \lra \wt \grg^*
\\
J_k^R : (g , \mu) & \lmt - \mu + kg^{-1}g^{\prime}, \tag  {1.13b}
\endalign
$$
respectively.  The functions obtained by pairing these maps with elements of
$\wt{\grg}$ Poisson commute with each other, but the maps are nonequivariant,
having as  $2-$cocycles $kc$ and $- kc$, respectively.
Thus, the Hamiltonian generators
$$
J_X^L := \langle J^L , X \rangle ,  \quad J_X^R := \langle J_k^R , X \rangle
\tag  {1.14}
$$
satisfy the following Poisson bracket relations:
$$
\align
\lbrace J_X^L , J_Y^L \rbrace & = J_{\lbrack X , Y \rbrack} ^L + kc(X , Y)
\tag
{1.15a}
\\
\lbrace J_X^R , J_Y^R \rbrace & = J_{\lbrack X , Y \rbrack} ^R - kc(X , Y)
\tag
{1.15b}
\\
\lbrace J_X^L , J_Y^R \rbrace & = 0  \ . \tag  {1.15c}
\endalign
$$
\endproclaim
\demo {Proof} Differentiating the actions (1.11), (1.12) along a
1-parameter subgroup $\lbrace h(t) = \text{exp}(-tX) \rbrace$, we find the
following representations of $\wt \grg$ in terms of functional vector fields
on $T^* \wt G$:
$$
\align
X_X^R = \langle gX , \frac {\delta}{\delta g} \rangle & - \langle \lbrack X ,
\mu \rbrack , \frac {\delta}{\delta \mu} \rangle  \tag  {1.16a}
\\
X_X^L = - \langle Xg , \frac {\delta}{\delta g} \rangle & - \langle k g^{-1}
X^{\prime} g , \frac {\delta}{\delta \mu} \rangle \ ,  \tag  {1.16b}
\\
\endalign
$$
where the first terms $\langle gX , \frac {\delta}{\delta g} \rangle$ and
$\langle Xg , \frac {\delta}{\delta g} \rangle$ denote, respectively, the
left and right invariant vector fields on the first factor $\wt G$ in
$T^* \wt G = \wt G \times \wt \grg^*$, with value $X$ at the identity
$e \in \wt G$, and the second terms, of the form $\langle \eta (g,\mu) , \frac
{\delta}{\delta \mu} \rangle$, denote the vector field on the second factor
$\wt \grg^*$ having value $\eta (g,\mu) \in \wt \grg^*$ at the point
$(g,\mu)$.  Evaluating the inner products with $\o_k$ gives
$$
\align
X_X^R  \bl \o_k & = \delta \langle \mu - kg^{-1}g^{\prime} ,X \rangle  \tag
{1.17a}
\\
X_X^L \bl \o_k & = - \delta \langle g \mu g^{-1} , X \rangle  \ ,
\tag {1.17b}
\endalign
$$
which shows that eqs\. (1.13a,b) give the moment maps generating these
actions.  To verify the Poisson bracket relations (1.15a-c), we compute:
$$
\align
\lbrace J_X^L , J_Y^L \rbrace = X_X^L (J_Y^L) & = \langle g \mu g^{-1} ,
\lbrack
X , Y \rbrack \rangle + k \langle X , Y^{\prime} \rangle
\\
\lbrace J_X^R , J_Y^R \rbrace = X_X^R (J_Y^R) & = \langle - \mu + k g^{-1}
g^{\prime} ,
\lbrack X , Y \rbrack \rangle - k \langle X , Y^{\prime} \rangle
\\
\lbrace J_X^L , J_Y^R \rbrace & = X_X^L (J_Y^R)  = 0\ .
\endalign
$$
\vskip-22pt
 \line{\hfill $\square$}
\enddemo

Proposition 1.1 implies that the map
$$
\align
J_k^{LR} : T^* \wt G & \lra (\wt \grg \oplus \wt \grg)^*
\\
J_k^{LR} : (g , \mu) & \lmt (g \mu g^{-1} , - \mu + kg^{-1} g^{\prime})
\tag {1.18}
\endalign
$$
to the dual of the direct sum  Lie algebra $\wt \grg \oplus \wt \grg$ is a
nonequivariant moment map generating the product of the left $(L)$ and right
$(R)$  translations, the nonequivariance given by $kc^{LR}$, where
$c^{LR}$ is the $2$-cocycle
$$
\align
c^{LR} : (\wt \grg \oplus \wt \grg) \times (\wt \grg \oplus \wt \grg) & \lra
\Bbb R
\\
c^{LR}((X_1 , Y_1) , (X_2 , Y_2)) & := \langle X_1 , X_2^{\prime} \rangle -
\langle
Y_1 , Y_2^{\prime} \rangle \ .  \tag  {1.19}
\endalign
$$

The centrally extended algebra $(\wt \grg \oplus \wt \grg)^{\wedge}$
associated with this cocycle is identified as the space
$\wt \grg + \wt \grg + \Bbb R$, with Lie
bracket
$$
\left[ (X_1 , Y_1 , a_1) , (X_2 , Y_2 , a_2) \right] = ( \lbrack X_1 , X_2
\rbrack , \lbrack Y_1 , Y_2 \rbrack , \langle X_1 , X_2^{\prime} \rangle -
\langle Y_1
, Y_2^{\prime} \rangle ) \ . \tag  {1.20}
$$
Again, $(\wt \grg \oplus \wt \grg)^\wedge$ is identified as a dense
subspace of the dual space $(\wt \grg \oplus \wt \grg)^{\wedge *}$ through
the pairing
$$
\align
\langle (U,V,a) , (X,Y,b) \rangle & := \langle U,X \rangle +
\langle V,Y \rangle + ab  \ , \tag {1.21}
\\
(U,V,a) & \in (\wt \grg \oplus \wt \grg)^{\wedge *} , (X,Y,b) \in (\wt \grg
\oplus \wt \grg)^\wedge.
\endalign
$$
The extended coadjoint representation on $(\wt \grg \oplus \wt
\grg)^{\wedge *}$ is given by the formulae
$$
\align
\wh{ad}_{(X,Y,b)}^*  (U,V,a) & = (ad_X^* U + aX^{\prime} , ad_Y^* V -
aY^{\prime} , 0) \tag {1.22a}
\\
\wh{Ad}_{(\wh{g,h})}^* (U,V,a) & = (Ad_g^* U + ag^{\prime}g^{-1}, Ad_h^* V -
ah^{\prime}h^{-1}, a)  \ ,   \tag{1.22b}
\endalign
$$
where again, the centrally extended group
$(\wt G \times \wt G)^\wedge \ra  \wt G \times \wt G$ acts through the
projection of any element  $(\wh{g,h}) \in (\wt G \times \wt G)^\wedge$ to its
 image $(g,h)$ in  $\wt G \times \wt G$.

{}From Proposition 1.1  we then have :
\proclaim
{Corollary 1.2}  The map
$$
\align
\hat {J}_k^{LR} : T^* \wt G & \lra (\wt \grg \oplus \wt \grg)^{\wedge *}
\\
\hat {J}_k^{LR} : (g, \mu) & \lmt (g \mu g^{-1} , - \mu + kg^{-1}g^{\prime} ,
k)  \tag
{1.23}
\endalign
$$
is an equivariant moment map with respect to the extended coadjoint action
(1.22a,b), generating the product action $L \times R$ of $\wt G \times \wt
G$ on $T^* \wt G$.
\endproclaim

The Lie Poisson bracket on $(\wt \grg \oplus \wt \grg)^{\wedge *}$ is
determined by:
$$
\align
\lbrace \langle \ .\  \vert (X_1,Y_1,b_1) \rangle ,& \langle \ .\  \vert
(X_2,Y_2,b_2) \rangle \rbrace \mid_{(U,V,a)}
\\
&= \langle U, \lbrack X_1, X_2 \rbrack \rangle + \langle V, \lbrack Y_1,Y_2
\rbrack \rangle
 + a( \langle X_1, X_2^{\prime} \rangle - \langle Y_1 , Y_2^{\prime} \rangle )
\tag  {1.24}
\endalign
$$
where $\langle \ .\  \vert (X,Y,b) \rangle$ again denotes the linear
functional on $(\wt \grg \oplus \wt \grg)^{\wedge *}$ corresponding to an
element $(X,Y,b) \in (\wt \grg \oplus \wt \grg)^\wedge$.  The moment map
$\hat {J}_k^{LR}$ is thus a Poisson map with respect to the Poisson brackets
(1.10a-c) associated to the noncanonical symplectic form $\o_k$.
\bigskip
\line{1c. \quad {\it  Hamiltonian Action of $\wt G \ltimes \wt \grg_A$ on
$T^* \wt G$}\hfil}
\medskip
Viewing $\wt \grg$ as an additive group, denoted $\wt \grg_A$, upon
which $\wt G$ acts via the adjoint representation, we may form the
semi-direct product $\wt G \ltimes \wt \grg_A$, with group multiplication
defined by
$$
\align
(\wt G \ltimes \wt \grg_A) \times (\wt G \ltimes \wt \grg_A) & \ra \wt G
\ltimes \wt \grg_A
\\
(g,X) \times (h,Y) & := (gh, X + g Yg^{-1})  \tag  {1.25}
\\
g,h \in \wt G , & \quad  X,Y \in \wt \grg_A \ .
\endalign
$$
The corresponding Lie algebra is the semi-direct sum $\wt \grg \dotplus \wt
\grg_A$, identified as a vector space with $\wt \grg + \wt \grg$, with Lie
bracket
$$
\left[ (X_1,Y_1) , (X_2,Y_2) \right] := ( [X_1,X_2 ],\
[ X_1,Y_2 ] -  [X_2,Y_1 ] )  \tag  {1.26}
$$
The dual space $(\wt \grg \dotplus\wt \grg_A)^*$ is again identified with $\wt
\grg + \wt \grg_A$, with dual pairing
$$
\align
\langle (U,V) , (X,Y) \rangle & := \langle V,X \rangle + \langle U, Y
\rangle  \tag  {1.27}
\\
(U,V)  & \in (\wt \grg \dotplus \wt \grg_A)^* ,
 \ (X,Y) \in \wt \grg \dotplus \wt \grg_A \ .
\endalign
$$
(Note the reversal of conventions relative to (1.21).) The coadjoint
representation of elements $(X,Y) \in \wt \grg
\dotplus \wt \grg_A$, $(g,Y) \in \wt G \ltimes \wt \grg_A$ in the algebra and
group is given by
$$
\align
ad_{(X,Y)}^* (U,V) & = ( \lbrack X,U \rbrack, \lbrack X,V \rbrack +
 \lbrack Y,U \rbrack ) \tag  {1.28a}
\\
Ad_{(g,Y)}^* (U,V) & = (gUg^{-1}, gVg^{-1} + \lbrack Y, gUg^{-1} \rbrack) \ .
\tag {1.28b}
\endalign
$$

The following action of the additive group $\wt \grg_A$ on $T^* \wt G$ is
easily verified to leave the canonical symplectic form  $\o_0$ invariant
$$
\align
A_X : (g,\mu) & \lmt (g,\mu + (g^{-1} X g)^{\prime}) \tag {1.29}
\\
X & \in \wt \grg_A \ .
\endalign
$$
In the remainder of this subsection only the canonical symplectic structure
on $T^* \wt G$ will be used.
\proclaim{Lemma 1.3}  The $\wt \grg_A$ - action (1.29) is Hamiltonian, and
is generated by the equivariant moment map $J^A : T^* \wt G \lra \grg_A^*$
defined by the formula
$$
J^A = g^{\prime} g^{-1}  \tag  {1.30}
$$
(where the dual pairing (1.2) is applied to $\wt \grg_A \sim \wt \grg ,\
\wt\grg_A^* \sim \wt \grg^*$.)
\endproclaim
\demo{Proof} Differentiating (1.29) along a 1-parameter subgroup
$\lbrace -tX \rbrace$ with respect to $t$ shows that the infinitesimal action
is generated by the vector field
$$
X_X^A = - \langle {(g^{-1} Xg)}^{\prime}, \frac {\delta}{\delta \mu} \rangle
\ .
\tag  {1.31}
$$
Taking the inner product with $\o_0$ gives
$$
\align
X_X^A \bl \o_0 & = -  \langle g^{-1} Xg , {(g^{-1} \delta g)}^{\prime} \rangle
\\
& = - \delta \langle X , g^{\prime} g^{-1} \rangle \ ,  \tag  {1.32}
\endalign
$$
showing that the flow is generated by the Hamiltonian $- J_X^A$, where
$$
J_X^A := \langle J^A , X \rangle  \ .  \tag   {1.33}
$$
The equivariance follows from the fact that, for any $X, Y \in \wt \grg_A$,
the functions $J_X^A , J_Y^A$ depend only on the first factor $g$ in $(g, \mu)
\in T^* \wt G$, and hence, by (1.10a) they Poisson commute.
\hfill $\square$
\enddemo

We may now compose this action with the  left translation action (1.12)
for $k = 0$,
$$
L_h : (g, \mu) \lmt (hg, \mu) \ ,  \tag  {1.34}
$$
to obtain an action $LA : (\wt G \ltimes \grg_A) \times T^* \wt G \lra T^* \wt
G$ of the semi-direct product group $\wt G \ltimes \wt \grg_A$ defined by:
$$
LA_{(h, X)} : (g, \mu) \lmt {(hg, \mu + (g^{-1} h^{-1} X h g)}^{\prime}). \tag
{1.35}
$$
It is easily verified that these maps compose correctly to define an action of
$\wt G \ltimes \wt \grg_A$ on $T^* \wt G$ and, by Lemma 1.3 and the fact
that the $L$-action (1.34) is Hamiltonian, the combined $LA$-action (1.35)
is  Hamiltonian as well.  However, the resulting moment map
$$
\align
J^{LA} : T^* \wt G & \lra {(\wt \grg \dotplus \wt \grg_A)}^*
\\
J^{LA} : (g, \mu) & \lmt (J^A (g, \mu), J^L (g, \mu) )
\\
&\quad  = (g^{\prime} g^{-1}, g \mu g^{-1} ) \tag  {1.36}
\endalign
$$
is no longer equivariant, as indicated in the following
\proclaim
{Proposition 1.4} The moment map $J^{LA}$ generating the $\wt G
\ltimes \wt \grg_A$ action is nonequi\-variant, having as cocycle:
$$
\align
c^{LA} : (\wt \grg \dotplus \wt \grg_A) \times (\wt \grg \dotplus \wt \grg_A) &
\lra
\Bbb R
\\
c^{LA}((X_1,Y_1) , (X_2, Y_2)) & := \langle X_1 , Y_2^{\prime} \rangle -
\langle
X_2 , Y_1^{\prime} \rangle  \ . \tag  {1.37}
\endalign
$$
Thus, the Hamiltonian generators
$$
\align
J_{(X,Y)}^{LA} & := \langle g \mu g^{-1}, X \rangle + \langle g^{\prime}
g^{-1}, Y
\rangle  \tag {1.38}
\\
X \in \wt \grg ,  &\quad  Y \in \wt \grg_A
\endalign
$$
satisfy the following Poisson bracket relations
$$
\lbrace J_{(X_1,Y_1)}^{LA} , J_{(X_2,Y_2)}^{LA} \rbrace =
J_{( [X_1,X_2 ], [ X_1,Y_2 ] - [ X_2,Y_1 ] )}^{LA} +
\langle X_1,Y_2^{\prime} \rangle - \langle X_2,Y_1^{\prime} \rangle  \ .
\tag{1.39}
$$
\endproclaim
\demo{Proof}  Eq\. (1.39) reduces to the following relations
$$
\align
\lbrace J_X^L,J_Y^L \rbrace & = J_{\lbrack X,Y \rbrack}^L  \tag {1.40a}
\\
\lbrace J_X^L,J_Y^A \rbrace & = J_{\lbrack X,Y \rbrack}^A + \langle
X,Y^{\prime}
\rangle \tag {1.40b}
\\
\lbrace J_X^A, J_Y^A \rbrace & = 0  \ . \tag{1.40c}
\endalign
$$
Eq\. (1.40a) is the particular case of eq\. (1.15a) with $k = 0$, while
eq\. (1.40c) is implied by Lemma 1.3.  Eq\.  (1.40b) is verified directly:
$$
\align
\lbrace J_X^L, J_Y^A \rbrace & = - X_Y^A (J_X^L)
\\
& = \langle (g^{-1} Y g)^{\prime} , \frac {\delta}{\delta \mu} \rangle \,
(\langle g \mu g^{-1} , X \rangle)
\\
& = \langle g^{\prime} g^{-1} , \lbrack X,Y \rbrack \rangle + \langle X,
Y^{\prime} \rangle ,
\endalign
$$
where eq\.  (1.31) was used in the second line.
  \hfill $\square$
\enddemo
The non-equivariance of the map $J^{LA} : T^* \wt G \ra {(\wt \grg \dotplus \wt
\grg_A)}^*$ with respect to the group action (1.35) may be expressed in
finite form as:
$$
J^{LA} \circ LA_{(h,X)} = Ad_{(h,X)}^* \circ J^{LA} + C^{LA} (h,X)\ , \
\tag{1.41}
$$
where the group 1-cocycle $C^{LA} : \wt G \ltimes \wt \grg_A \ra (\wt \grg
\dotplus \grg_A)^*$ is defined by the formula:
$$
C^{LA} (h,X) := ( h^{\prime} h^{-1}, X^{\prime} -
\lbrack h^{\prime} h^{-1}, X \rbrack)  \ .\tag {1.42}
$$
The cocycle relation
$$
C^{LA}(gh, X + gYg^{-1}) = Ad_{(g,X)}^* C^{LA} (h,Y) + C^{LA} (g,X)
 \tag  {1.43}
$$
is verified by applying the $Ad^*$-action (1.28b) to the definition (1.42).

Using the algebra $2-$cocycle $c^{LA}$ given in eq\. (1.37), we define the
centrally extended Lie algebra
$(\wt \grg \dotplus  \wt \grg_A)^\wedge$ as the space $\wt \grg + \wt \grg +
\Bbb
R$, with Lie bracket
$$
\lbrack (X_1,Y_1,a_1), (X_2,Y_2,a_2) \rbrack = ( \lbrack X_1,X_2 \rbrack,
\lbrack X_1,Y_2 \rbrack - \lbrack X_2,Y_1 \rbrack, \langle X_1,Y_2^{\prime}
\rangle - \langle X_2,Y_1^{\prime} \rangle ) \ . \tag  {1.44}
$$
A dense subspace of the dual $(\wt \grg \dotplus \wt \grg_A)^{\wedge *}$ is
again identified with $(\wt \grg \dotplus \wt \grg_A)^\wedge$, with typical
elements denoted again $(U, V, a)$. The notational conventions are such that
the pairing of eq\. (1.21) is replaced by:
$$
\align
\langle (U,V,a) , (X,Y,b) \rangle_A & := \langle V,X \rangle +
\langle U,Y \rangle + ab , \tag {1.21$_A$}
\\
(U,V,a) \in (\wt \grg \oplus \wt \grg_A)^{\wedge *} &,
\quad(X,Y,b) \in (\wt \grg
\oplus \wt \grg_A)^\wedge.
\endalign
$$

The extended coadjoint representation on $(\wt \grg \dotplus \wt
\grg_A)^{\wedge *}$ is then given by the formulae:
$$
\align
\wh{ad}_{(X,Y,a)}^* (U,V,b) & =
(\lbrack X, U \rbrack + bX^{\prime}, \lbrack X,V \rbrack + \lbrack Y, U \rbrack
+ bY^{\prime}, 0)  \tag  {1.45a}
\\
\wh{Ad}_{(\wh{g,Y})}^* (U,V,b)  & = ( gUg^{-1} + bg^{\prime}g^{-1}, gVg^{-1} +
\lbrack Y, gUg^{-1} \rbrack + bY^{\prime} -
b \lbrack g^{\prime}g^{-1}, Y \rbrack, b )  \ . \tag{1.45b}
\endalign
$$
{}From Proposition 1.4 then follows
\proclaim
{Corollary 1.5}  The map
$$
\align
\hat J^{LA} : T^* \wt G & \lra (\wt \grg \oplus \wt \grg_A)^{\wedge *}
\\
\hat J^{LA} : (g, \mu) & \lmt (g^{\prime} g^{-1}, g \mu g^{-1}  , 1)
\tag{1.46}
\endalign
$$
is an equivariant moment map with respect to the extended coadjoint action
(1.45a,b), generating the action $LA : (\wt G \ltimes \wt \grg_A) \times T^*
\wt G \ra T^* \wt G$ defined by (1.35).
\endproclaim
The Lie-Poisson bracket on $(\wt \grg \dotplus \wt \grg_A)^{\wedge *}$ is
determined by the formula:
$$
\align
&\lbrace \langle \ .\  \vert (X_1,Y_1,b_1) \rangle_A , \langle \ .\  \vert
(X_2,Y_2,b_2) \rangle_A \mid_{(U,V,a)}\\
&\qquad = \langle V, \lbrack X_1,X_2 \rbrack \rangle + \langle U, \lbrack
X_1,Y_2 \rbrack - \lbrack X_2,Y_1 \rbrack \rangle + a( \langle X_1,Y_2^{\prime}
\rangle - \langle X_2,Y_1^{\prime} \rangle )   \ . \tag  {1.47}
\endalign
$$
\smskip
The moment map $\hat J^{LA}$ is thus a Poisson map with respect to the
canonical Poisson brackets on $T^* \wt G$ given by setting $k = 0$ in
eqs\. (1.10a-c).
\bigskip
\line{1d. \quad{\it  Hamiltonian Action of $\Cal D_1 \ltimes \Cal D_0$}\hfil}
\medskip
Let $\Cal D_1 := \text {Diff } S^1$ (resp. Diff $\Bbb R)$ be the
group of diffeomorphisms of $S^1$  (resp. $\Bbb R)$ and $\Cal D_0
:= {\Cal F}^{\times} (S^1)$ (resp. $\Cal F^\times (\Bbb R)$) the space
of smooth non-vanishing functions on $S^1$ (resp. $\Bbb R$), viewed as
an abelian group under multiplication.  Using the natural action of
$\Cal D_1$ on $\Cal D_0$
$$
\align
\wt \sigma : \Cal D_0 & \ra \Cal D_0
\\
\wt \sigma :  f & \lmt f \circ  \wt{\sigma}^{-1}  \tag  {1.48}\\
\wt \sigma & \in \Cal D_1 \ ,
\endalign
$$
we define the semi-direct product $\Cal D_1 \ltimes \Cal D_0$, with group
multiplication
$$
\align
(\Cal D_1 \ltimes \Cal D_0) \times (\Cal D_1 \ltimes \Cal D_0) & \lra \Cal D_1
\ltimes \Cal D_0
 \\
(\wt \sigma, f_1) \times (\wt \sigma_2, \wt  f_2)
& \lmt (\wt
\sigma_1 \circ \wt \sigma_2,  f_1 \ f_2 \circ \wt \sigma_1^{-1})
\tag  {1.49}\\
\wt \sigma_1, \wt \sigma_2 \in \Cal D_1 , & \quad f_1,  f_2 \in \Cal D_0 \ .
\endalign
$$

The natural action of $\Cal D_1 \ltimes \Cal D_0$ on the space $\Cal F$ of
smooth functions on $S^1$ (resp. $\Bbb R$) is:
$$
\align
(\wt \sigma,  f) : \Cal F & \lra \Cal F
\\
(\wt \sigma,  f) : h & \lmt  f \ h \circ \wt \sigma^{-1} .  \tag{1.50}
\endalign
$$
The Lie algebra of $\Cal D_1 \ltimes \Cal D_0$ may be identified with  the
space $\text{diff}_1$ of differential operators of order $\leq 1$, with
typical elements $\alpha \frac {d}{d\sigma} + \beta \in \text {diff}_1$
denoted by pairs $(\alpha,\beta) \in \Cal F \oplus \Cal F$, and the Lie
product given by commutation:
$$
\lbrack (\alpha_1,\beta_1), (\alpha_2,\beta_2) \rbrack = (\alpha_1
\alpha_2^{\prime} - \alpha_2 \alpha_1^{\prime} , \alpha_1 \beta_2^{\prime} -
\alpha_2 \beta_1^{\prime})  \ .
\tag  {1.51}
$$
The dual space $\text {diff}_1^{\ *}$ is identified with  the space
$\Cal F \oplus \Cal F$
through the dual pairing
$$
\align
\text {diff}_1^{\ *}\times \text {diff}_1 & \ra \Bbb R
\\
\langle (v, w) , (\alpha,\beta) \rangle & := \langle v \alpha \rangle +
\langle w, \beta \rangle  \tag  {1.52}
\\
(v, w) \in \text {diff}_1^{\ *}, & \quad (\alpha, \beta) \in \text {diff}_1 \ ,
\endalign
$$
where the integrals
$$
\langle v, \alpha \rangle := \int v(\sigma)\alpha(\sigma) d
\sigma ,  \qquad \langle w,\beta \rangle := \int w (\sigma)
\beta(\sigma) d \sigma  \tag  {1.53}
$$
are taken over the appropriate domain ($S^1$ or $\Bbb R$), and the
functions $\alpha, \beta, u, v \in \Cal F$ are assumed  to be
integrable as required.

Let
$$
\align
\Theta : {\Bbb R}^{\times} \times {\Bbb R}^{\times} & \lra G  \tag  {1.54a}
\\
\Phi : {\Bbb R}^{\times} \times {\Bbb R}^{\times} & \lra G  \tag  {1.54b}
\\
\Theta (a,b) := \Theta_{0} (a) \Theta_{1} (b),
& \qquad \Phi (a,b) := \Phi_{0} (a) \Phi_{1}
(b)  \tag  {1.54c} \\
a,b  & \in {\Bbb R}^{\times}
\endalign
$$
be a pair of homomorphisms into $G$ from the direct product group
 ${\Bbb R}^{\times} \times {\Bbb R}^{\times}$ formed from two copies of
the multiplicative group  $\Bbb R^{\times}$ of non-zero reals. The
derivatives $\Theta_*\vert _{(1,1)} \in \text {Hom} (\Bbb R^2 ,
\grg), \,  \Phi_* \vert_{(1,1)} \in \text {Hom} (\Bbb R^2 , \grg)$
at  the identity element $(1,1)$ may be expressed
$$
\Theta_*\vert_{(1,1)} (x,y) = x \theta_0 + y \theta_1, \qquad
\Phi_*\vert_{(1,1)} (x,y) = x \phi_0 + y \phi_1   \ , \tag  {1.55}
$$
where the elements $\theta_0, \theta_1, \phi_0, \phi_1 \in \grg$ satisfy
$$
\lbrack \theta_0 , \theta_1 \rbrack = 0  , \qquad
 \lbrack \phi_0 , \phi_1 \rbrack = 0   \ . \tag  {1.56}
$$

We define a class of right $\Cal D_1 \ltimes \Cal D_0$ actions on
$T^* \wt G$, parametrized by $(k , \Theta , \Phi)$:
$$
\align
(\wt \sigma , f) : T^* \wt G & \lra T^* \wt G
\\
(\wt \sigma , f) : (g , \mu) & \lmt (\check g , \check\mu)  \tag  {1.57a}
\endalign
$$
where
$$
\align
\check g & := \Theta_1 (\wt \sigma^{\prime}) \Theta_0 (f \circ \wt \sigma) (g
\circ \wt
\sigma) \Phi_0 (f \circ \wt \sigma)^{-1} \Phi_1 (\wt \sigma^{\prime})^{-1}
\tag  {1.57b}
\\
\check\mu & := \wt \sigma^{\prime} \Phi_1 (\wt \sigma^{\prime}) \Phi_0 (f \circ
\wt \sigma)
(\mu \circ \wt \sigma) \Phi_0 (f \circ \wt \sigma)^{-1} \Phi_1 (\wt
\sigma^{\prime})^{-1}
\\
& + k \wt \sigma^{\prime} (f^{\prime} \circ \wt \sigma) \Phi_1 (\wt
\sigma^{\prime}) \Phi_0 (f \circ \wt
\sigma) (g \circ \wt \sigma)^{-1} \theta_0 (g \circ \wt \sigma) \Phi_0 (f \circ
\wt
\sigma)^{-1} \Phi_1 (\wt \sigma^{\prime})^{-1}
\\
& + k \frac {\wt \sigma^{\prime \prime}}{\wt \sigma^{\prime}} \Phi_1 (\wt
\sigma^{\prime})
\Phi_0 (f \circ \wt \sigma) (g \circ \wt \sigma)^{-1} \theta_1 (g \circ \wt
\sigma)\Phi_0 (f \circ \wt \sigma)^{-1} \Phi_1 (\wt \sigma^{\prime})^{-1}  \ .
\tag  {1.57c}
\endalign
$$
Although this action appears complicated at first sight, it is actually
very simple when decomposed as a product of the $\Cal D_0$-action:
$$
(1,f) : (g, \mu) \lmt (\Theta_0 (f) g \Phi_0 (f)^{-1},  \Phi_0 (f) \mu \Phi_0
(f)^{-1} + k f^{\prime} \Phi_0 (f) g^{-1} \theta_0 g \Phi_0 (f)^{-1})
\tag  {1.58}
$$
and the $\Cal D_1$-action:
$$
(\wt \sigma , 1)  : (g , \mu) \lra  (\check g , \check\mu)  \tag  {1.59a}
$$
defined by
$$
\align
\check g  =& \Theta_1 (\wt \sigma^{\prime}) (g \circ \wt \sigma) \Phi_1 (\wt
\sigma^{\prime})^{-1}  \tag  {1.59b}
\\
\check \mu  = &\wt \sigma^{\prime} \Phi_1 (\wt \sigma^{\prime}) (\mu \circ \wt
\sigma) \Phi_1 (\wt
\sigma^{\prime})^{-1}
 + k \frac {\wt \sigma^{\prime \prime}}{\wt \sigma^{\prime}} \Phi_1 (\wt
\sigma^{\prime}) (g \circ \wt
\sigma)^{-1} \theta_1 (g \circ \wt \sigma) \Phi_1 (\wt \sigma^{\prime})^{-1}
\tag
{1.59c}
\endalign
$$

The $\Cal D_0$-action (1.58) is just left translation (1.12) by the element
$\Theta_0 (f) \in \wt G$ and right translation (1.11) by the element
$\Phi_0 (f) \in \wt G$.  The $\Cal D_1$-action (1.59a-c) is the cotangent
bundle lift of the natural right action
$$
\align
\wt \sigma : \wt G & \lra \wt G
\\
\wt \sigma : g & \lmt g \circ \wt \sigma,  \tag  {1.60}
\endalign
$$
composed with a ``twisting'' by the left translation action (1.12) of
$\Theta_1 (\wt \sigma^{\prime}) \in \wt G$ and right translation (1.11) by
$\Phi_1 (\wt \sigma^{\prime}) \in \wt G$.  It is verified directly that the
maps
(1.57a-c) do indeed compose correctly to define a right $\Cal D_1 \ltimes
\Cal D_0$ action. Moreover, by considering the separate maps that are
composed to form (1.57a-c), it is easily verified that this action preserves
the symplectic form $\o_k$ on $T^* \wt G$.  The infinitesimal action
obtained from (1.57a-c) by differentiating the flow given by the
$1-$parameter group
$\lbrace \text {exp}[-t (\alpha , \beta)] \subset \Cal D_1 \ltimes \Cal D_0
\rbrace$ at $t = 0$ is represented by the functional vector
fields:
$$
X_{(\alpha,\beta)}^{\Cal {D}}  = X_{\alpha}^1 + X_{\beta}^0  \tag  {1.61a}
$$
where
$$
X_{\beta}^0 :=  X_{\beta \theta_0}^L + X_{\beta \phi_0}^R
\tag{1.61b}
$$
generates the $\Cal D_0$-action (1.58), and
$$
X_{\alpha}^1  :=  - \langle \alpha g^{\prime} , \frac {\delta}{\delta g}
\rangle -
\langle (\alpha \mu)^{\prime} , \frac {\delta}{\delta \mu} \rangle\ +
X_{\alpha^{\prime} \theta_1}^L + X_{\alpha^{\prime} \phi_1}^R
 \tag {1.61c}
$$
generates the $\Cal D_1$-action (1.59a-c).
(In the above, we have chosen signs so that the map $(\alpha , \beta) \ra
X_{(\alpha , \beta)}^{\Cal D}$ defines a homomorphism to the Lie algebra of
functional vector fields on $T^* \wt G$, rather than an anti-homomorphism.
This means the flows generate the left action of $\Cal D_1 \ltimes \Cal D_0$
defined by replacing $(\wt \sigma , f)$ in (1.57a-c) by the inverse $(\wt
\sigma^{-1} , (f \circ \wt \sigma)^{-1})$).
The Hamiltonian properties of this action are summarized in the following:
\proclaim
{Proposition 1.6}  The action (1.57a-c) is Hamiltonian, with the
1-parameter subgroups $\lbrace \text{exp}[-t (\alpha , \beta)] \rbrace$
generated as the flow of the Hamiltonians:
$$
\align
J_{(\alpha , \beta)}^{\Cal D}  := & \langle J^L , \alpha^{\prime} \theta_1
\rangle
+ \langle J^R , \alpha' \phi_1 \rangle
\\
& \ + \frac {1}{2k} ( \langle \alpha J^L , J^L \rangle - \langle \alpha J^R ,
J^R
\rangle
\\
&\ + \langle J^L, \beta \theta_0 \rangle + \langle J^R, \beta \phi_0 \rangle
\ . \tag {1.62}
\endalign
$$
Thus, the map
$$
\align
J^{\Cal D} : T^* \wt G & \lra \text {diff}_1^{\ *}
\\
J^{\Cal D} : (g , \mu) & \lmt (J^0 , J^1)  \tag  {1.63a}
\endalign
$$
defined by
$$
\align
J^0  =& \Cal{B}(J^L , \theta_0) + \Cal{B}(J^R , \phi_0)  \tag  {1.63b}
\\
J^{1}  =& - \Cal{B} (J^{L{\prime}} , \theta_1) - \Cal{B} (J^{R{\prime}} ,
\phi_1)
\\
&\  + \frac {1}{2k} (\Cal{B} (J^L , J^L) - \Cal{B} (J^R , J^R))  \tag {1.63c}
\endalign
$$
is the moment map generating this action.  It is nonequivariant, satisfying
the Poisson bracket relations:
$$
\align
\lbrace J_{(\alpha_1 , \beta_1)}^{\Cal D} , J_{(\alpha_2 , \beta_2)}^{\Cal D}
\rbrace  = &J_{\lbrack \alpha_1 \alpha_2^{\prime} - \alpha_2 \alpha_1^{\prime}
, \alpha_1
\beta_2^{\prime} - \alpha_2 \beta_1^{\prime} \rbrack}^{\Cal D}
\\
&\ + k \lbrack \Cal{B}(\theta_0 , \theta_0) - \Cal{B}(\phi_0 , \phi_0) \rbrack
\langle
\beta_1 ,\beta_2^{\prime} \rangle
\\
& \ + k \lbrack \Cal{B}(\theta_1 , \theta_0) - \Cal{B}(\phi_1 , \phi_0) \rbrack
[\langle \alpha_1^{\prime}, \beta_2^{\prime}\rangle -
\langle \alpha_2^{\prime} ,\beta_1^{\prime} \rangle]
\\
& \ + k \lbrack \Cal{B}(\theta_1 , \theta_1) - \Cal{B}(\phi_1 , \phi_1)
 \rbrack \langle
\alpha_1^{\prime}, \alpha_2^{\prime \prime} \rangle   \ .\tag  {1.64}
\endalign
$$
\endproclaim
\demo
{Proof}  Let
$$
\align
J_{\alpha}^1 & := \langle J^1 , \alpha \rangle = J_{(\alpha,0)}^{\Cal D}
 \tag {1.65a}
\\
J_{\beta}^0 & := \langle J^0 , \beta \rangle = J_{(0 , \beta)}^{\Cal D}  \ .
\tag{1.65b}
\endalign
$$
To show that $J^0$ and $J^1$ generate the ${\Cal D}_0$ and ${\Cal D}_1$
actions respectively, we must verify
$$
\align
X_{\alpha}^1 \bl \o_k & = - \delta J_{\alpha}^1 \tag {1.66a}
\\
X_{\beta}^0 \bl \o_k & = - \delta J_{\beta}^0 \ .  \tag {1.66b}
\endalign
$$
$J_{\alpha}^1$ and $J_{\beta}^0$ may be expressed in terms of the left and
right moment maps $J^L , J^R$ as:
$$
\align
J_{\alpha}^1 & = J_{\alpha^ \prime {\theta_1}}^L + J_{\alpha ^\prime
\phi_1}^R + \frac{1}{2k} ( \langle \alpha J^L , J^L \rangle - \langle \alpha
J^R , J^R \rangle ) \tag  {1.67a}
\\
J_{\beta}^0 & = J_{\beta \theta_0}^L + J_{\beta \phi_0}^R \ .  \tag {1.67b}
\endalign
$$
Eq\. (1.66b) is immediate from the form (1.61b) of $X_{\beta}^0$ and eqs\.
(1.17a,b).  Similarly, using the form (1.61c) of $X_{\alpha}^1$, eq (1.66a)
reduces to:
$$
- \lbrack \langle \alpha g^{\prime}, \frac{\delta}{\delta g} \rangle +
\langle (\alpha  \mu)^\prime ,
\frac{\delta}{\delta \mu} \rangle \rbrack \bl
\o_k = \frac{1}{2k} \delta \lbrack \langle \alpha J^L , J^L \rangle - \langle
\alpha J^R , J^R \rangle \rbrack \  ,  \tag {1.68}
$$
which is verified directly.  The Poisson brackets (1.64) are equivalent to the
relations:
$$
\align
\lbrace J_{\alpha_1}^1 , J_{\alpha_2}^1 \rbrace & = J_{(\alpha_1
\alpha_2^{\prime} -\alpha_2 \alpha_1^{\prime} \rangle}
+k[\Cal B (\theta_1,\theta_1)  -  \Cal B(\phi_1,\phi_1)]
\langle \alpha_1', \alpha_2'\rangle
 \tag {1.68a}
\\
\lbrace J_{\alpha}^1 , J_{\beta}^0 \rbrace & = J_{\alpha \beta^{\prime}}^0 + k
\lbrack \Cal{B}(\theta_1 , \theta_0) - \Cal{B}(\phi_1 , \phi_0) \rbrack \langle
\alpha^{\prime} ,\beta^{\prime} \rangle \tag  {1.68b}
\\
\lbrace J_{\beta_1}^0 , J_{\beta_2}^0 \rbrace & = k \lbrack \Cal{B}(\theta_0 ,
\theta_0) - \Cal{B}(\phi_0 , \phi_0) \rbrack \langle \beta_1 , \beta_2^{\prime}
\rangle  \ . \tag  {1.68c}
\endalign
$$
This is verified using the decompositions (1.63b,c) and the Poisson bracket
relations
$$
\align
\lbrace J_X^L, \langle \alpha J^L , J^L \rangle \rbrace & = - 2kJ_{\alpha
X^{\prime}}^L  \tag  {1.69a}
\\
\lbrace J_X^R , \langle \alpha J^R , J^R \rangle \rbrace & = 2k J_{\alpha
X^{\prime}}^R  \tag  {1.69b}
\\
\lbrace \langle \alpha_1 J^L , J^L \rangle , \langle \alpha_2 J^L , J^L \rangle
\rbrace & = 2k \langle (\alpha_1 \alpha_2^{\prime} - \alpha_2
\alpha_1^{\prime} ) J^L , J^L \rangle  \tag  {1.69c}
\\
\lbrace \langle \alpha_1 J^R , J^R \rangle, \langle \alpha_2 J^R , J^R \rangle
\rbrace & = - 2k \langle (\alpha_1 \alpha_2^{\prime} - \alpha_2
\alpha_1^{\prime} ) J^R , J^R \rangle  \ , \tag  {1.69d}
\endalign
$$
which follow from (1.15a,b) and the Leibnitz rule.        \hfill $\square$
\enddemo
In the Poisson brackets (1.64) we may identify three 2-cocycles for the
algebra $\text{diff}_1$
$$
\align
c_i^{\Cal D} : \text{diff}_1  \times \text{diff}_1  &\ra \Bbb R ,
\qquad i = 1, 2, 3
\\
c_1^{\Cal D} ((\alpha_1 , \beta_1) , (\alpha_2 , \beta_2)) & := \langle \beta_1
, \beta_2^{\prime} \rangle  \tag  {1.70a}
\\
c_2^{\Cal D} ((\alpha_1 , \beta_1) , (\alpha_2 , \beta_2)) & := \langle
\alpha_1^{\prime} , \beta_2^{\prime} \rangle - \langle \alpha_2^{\prime} ,
\beta_1^{\prime} \rangle  \tag  {1.70b}
\\
c_3^{\Cal D} ((\alpha_1 , \beta_1) , (\alpha_2 , \beta_2)) & := \langle
\alpha_1^{\prime} , \alpha_2^{\prime \prime} \rangle  \ .  \tag{1.70c}
\endalign
$$
Any combination of these may be used to define central extensions of the
algebra.  In particular, define the cocycle:
$$
c^{\Cal D} := l_1 c_1^{\Cal D} + l_2 c_2^{\Cal D} + l_3 c_3^{\Cal D}  \ ,
 \tag
{1.71}
$$
where
$$
\align
l_1 & := \Cal{B}(\theta_0 , \theta_0) - \Cal{B}(\phi_0 , \phi_0)  \tag{1.72a}
\\
l_2 & := \Cal{B}(\theta_1 , \theta _0) - \Cal{B}(\phi_1, \phi_0)  \tag{1.72b}
\\
l_3 & := \Cal{B}(\theta_1 , \theta_1) - \Cal{B}(\phi_1 , \phi_1)  \ .
 \tag{1.72c}
\endalign
$$
The centrally extended algebra $\text{diff}_1^{\ \wedge (l_1 l_2 l_3)}$
associated with this
cocycle is identified with the space $\text{diff}_1 \oplus \Bbb R$, with Lie
brackets
$$
\align
\lbrack (\alpha_1 , \beta_1 , a_1) , (\alpha_2 , \beta_2 , a_2)\rbrack &=
(\alpha_1 \alpha_2^{\prime} - \alpha_2 \alpha_1^{\prime} , \alpha_1
\beta_2^{\prime} - \alpha_2 \beta_1^{\prime} ,\\
 & \quad l_1 \langle \beta_1 , \beta_2^{\prime} \rangle + l_2 (\langle
\alpha_1^{\prime} , \beta_2^{\prime} \rangle - \langle \alpha_2^{\prime} ,
\beta_1^{\prime} \rangle )
+ l_3 \langle \alpha_1^{\prime} ,\alpha_2^{\prime \prime} \rangle )
\tag {1.73}
\\
(\alpha_1 , \beta_1) , (\alpha_2 , \beta_2) & \in \text{diff}_1  ,
\qquad a_1 , a_2 \in  \Bbb R  \ .
\endalign
$$
A dense subspace of the dual space $\text{diff}_1^{\ \wedge (l_1 l_2 l_3) *}$
is as usual
identified with $\text{diff}_1 \oplus \Bbb R$, through the pairing:
$$
{\langle (v,w,a) , (\alpha, \beta, b)\rangle}_{\Cal D}   := \langle
v, \alpha \rangle + \langle w, \beta \rangle + ab   \ . \tag{1.74}
$$
The extended coadjoint representation of
$\text{diff}_1^{\ \wedge (l_1 l_2 l_3)}$ on
$\text{diff}_1^{\ \wedge (l_1 l_2 l_3) *}$ is given by the formula
$$
\wh{ad}_{(\alpha, \beta, b)}^{*} (v,w,a) = (2 \alpha^{\prime} v + \alpha
v^{\prime} + w \beta^{\prime} - a l_2 \beta^{\prime \prime} - a l_3
\alpha^{\prime \prime \prime} , \alpha^{\prime} w + \alpha w^{\prime} + a
l_1 \beta^{\prime} + a l_2 \alpha^{\prime \prime}, 0)  . \tag  {1.75}
$$
It follows from Proposition 1.6 that we may define an extended,
equivariant moment map  from $T^{*} \wt G$ to
$\text{diff}_1^{\ \wedge (l_1 l_2 l_3) *}$ that generates the same
$\Cal D_1 \ltimes \Cal D_0$ action (1.57a-c).
\proclaim
{Corollary 1.7} The map
$$
\align
\hat J^{\Cal D} : T^{*} \wt G & \lra \text{diff}_1^{\ \wedge (l_1 l_2 l_3) *}
\sim
\text{diff}_1^{\ *} \oplus \Bbb R
\\
\hat J^{\Cal D} : (g , \mu) & \lmt (J^{\Cal D} , k)  \tag  {1.76}
\endalign
$$
is an equivariant moment map with respect to the extended coadjoint action
(1.75), generating the $\Cal D_1 \ltimes \Cal D_0$ action (1.57a-c).
\endproclaim
The Lie Poisson bracket on $\text{diff}_1^{\ \wedge (l_1 l_2 l_3) *}$ is
determined by
the formula:
$$
\align
\lbrace {\langle \ .\ \vert (\alpha_1 , \beta_1 , b_1) \rangle}_{\Cal D} &,
{\langle \ .\ \vert (\alpha_2 , \beta_2 , b_2) \rangle}_{\Cal D} \rbrace
\mid_{(v,w,a)}
\\
& = \langle v, \alpha_1 \alpha_2^{\prime} - \alpha_2 \alpha_1^{\prime} \rangle
+
\langle w, \alpha_1 \beta_2^{\prime} - \alpha_2 \beta_1^{\prime} \rangle +
al_1 \langle \beta_1 , \beta_2^{\prime} \rangle \\
& \qquad + al_2 (\langle \alpha_1^{\prime} , \beta_2^{\prime} \rangle -
\langle \alpha_2^{\prime} ,
\beta_1^{\prime} \rangle ) + al_3 \langle \alpha_1^{\prime} , \alpha_2^{\prime
\prime} \rangle  \ ,
\tag {1.77}
\endalign
$$
where, as usual, ${\langle \ .\ \vert (\alpha, \beta, b) \rangle}_{\Cal D}$
denotes the linear functional on $\text{diff}_1^{\ \wedge (l_1 l_2 l_3) *}$
associated by dual
pairing to the algebra element $(\alpha, \beta, b) \in
\text{diff}_1^{\ \wedge (l_1 l_2 l_3)}$. The map (1.76) is thus a Poisson map
with respect
to the Poisson brackets (1.10a-c) on $T^{*} G$ associated to the symplectic
form $\o_k$, and the Lie Poisson brackets (1.77) on
$\text{diff}_1^{\ \wedge (l_1 l_2 l_3)*}$.

We mention at this point that for the applications to integrable systems to
be discussed in section 2, a particular case of the $\Cal D_1 \ltimes \Cal D_0$
group action (1.57a-c) will be considered; namely, when the elements
$\theta_0,\ \theta_1,\  \phi_0,\  \phi_1  \in \gr g$ are of the form
$$
\theta_0 = -2\theta_1 = -2\phi_1 =: 2 \theta , \quad \phi_0 = 0 \ .
\tag  {1.78a}
$$
Hence
$$
l_1 = -2 l_2 = 4\Cal{B}(\theta, \theta)=: 4l ,  \quad  l_3 = 0 \ ,
 \tag  {1.78b}
$$
and the cocycle $c^{\Cal D}$ takes the form
$$
c^{\Cal D} ((\alpha_1 , \beta_1) , (\alpha_2 , \beta_2)) =
 l( 4\langle \beta_1 , \beta_2^{\prime} \rangle -
2\langle \alpha_1^{\prime} , \beta_2^{\prime} \rangle +
2\langle \alpha_2^{\prime}  , \beta_1^{\prime} \rangle )  \ .
\tag  {1.78c}
$$
The Poisson brackets (1.77) thus reduce to (cf\.  eq\. (2.36)):
$$
\align
\lbrace {\langle  \ . \ \vert (\alpha_1 , \beta_1, b_1) \rangle}_{\Cal D} ,
{\langle \ . \ \vert (\alpha_2 , \beta_2, b_2) \rangle}_{\Cal D}
\mid_{(v,w,a)} &
\\
= \langle v, \alpha_1 \alpha_2^{\prime} - \alpha_2 \alpha_1^{\prime} \rangle +
\langle w ,
\alpha_1 \beta_2^{\prime} - \alpha_2 \beta_1^{\prime} \rangle +
4al \langle \beta_1 , \beta_2^{\prime} \rangle - &
2a l ( \langle \alpha_1^{\prime} , \beta_2^{\prime} \rangle -
\langle\alpha_2^{\prime} , \beta_1^{\prime}
\rangle )   \ . \tag  {1.79}
\endalign
$$

Note that for $k \neq 0$ the maps $J^{\Cal D} : T^{*} \wt G \ra
\text{diff}_1^{\ *}, \quad  \hat J^{\Cal D} : T^{*} \wt G \ra
\text{diff}_1^{\ \wedge (l_1 l_2 l_3)*}$
 defined by  eqs\. (1.63a-c), (1.76) factor through the map
     $J^{LR}: T^{*} \wt G \ra {(\wt {\gr g} \oplus \wt {\gr g})}^{*}$ or
${\hat J}^{LR}:T^{*} \wt G \ra {(\wt {\gr g }\oplus \wt {\gr g})}^{\wedge *}$,
suggesting that the $\Cal D_1 \ltimes \Cal D_0$ group action (1.57a-c) also
induces a Hamiltonian action on ${(\wt {\gr g} \oplus \wt {\gr g})}^{\wedge
*}$.  This is in fact the case, and the action
is easily computed.
\proclaim
{Proposition 1.8}  The moment map
${\hat J}^{LR}:T^{*} \wt G \ra {(\wt {\gr g} \oplus \wt {\gr g})}^{\wedge *}$
is equivariant with respect to the $\Cal D_1 \ltimes \Cal D_0$ action
(1.57a-c) on $T^{*} \wt G$ and the following action on ${(\wt \grg \oplus \wt
\grg)^{\wedge *}}$:
$$
\align
(\wt \sigma , f) : {(\grg \oplus \wt {\grg})}^{\wedge *} & \ra {(\wt {\grg}
\oplus \wt {\grg})}^{\wedge *}
\\
(\wt \sigma , f) : (U,V,a) & \ra (\check U, \check V, a)  \tag  {1.80a}
\endalign
$$
where
$$
\align
\check U = &\wt{\sigma}'\Theta_1 (\wt{\s}')\Theta_0(f\circ \wt{\s})
(U\circ \wt{\s})\Theta_0(f\circ \wt{\s})^{-1}
\Theta_1(\wt{\s}')^{-1} + a (f\circ \wt{\s})'\theta_0
 + a \frac{\wt{\s}''}{\wt{\s}'}\theta_1   \tag{1.80b}\\
\check V  = &\wt{\sigma}'\Phi_1 (\wt{\s}')\Phi_0(f\circ \wt{\s})
(V\circ \wt{\s}) \Phi_0(f\circ \wt{\s})^{-1}
\Phi_1(\wt{\s}')^{-1} - a (f\circ \wt{\s})'\phi_0 -
a \frac{\wt{\s}''}{\wt{\s}'}\phi_1   \ .  \tag{1.80c}
\endalign
$$
\endproclaim
\demo{Proof}  This is verified by directly substituting the RHS of
eqs\. (1.57b,c) into the definitions (1.13a,b) of $J^L$ and $J^R$. \hfill
$\square$
\enddemo

The Hamiltonian properties of this action are summarized in the following.
\proclaim
{Theorem 1.9}  For  $k \neq 0$, the moment map $\hat J^{\Cal D}$ factors
into:
$$
\hat J^{\Cal D}  = \hat J^d  \circ  \hat J^{LR}         \tag {1.81}
$$
where the map  $\hat J^d : (\wt \grg \oplus \wt {\grg})^{\wedge *}  \lra
\text{diff}_1^{\ \wedge (l_1 l_2 l_3) *}$, defined by the formulae
$$
\align
\hat J^d (U,V,a) &= (J^{d1}, J^{d0}, a)      \tag {1.82a}
\\
J^{d 1} (U,V,a) &:=  - \Cal{B}(U^{\prime }, \theta_1) - \Cal{B}(V^{\prime }
,\phi_1) + \frac
{1} {2a} (\Cal{B}(U,U) - \Cal{B}(V,V))    \tag  {1.82b}
\\
J^{d0} (U,V,a) &:= \Cal{B}(U,\theta_0) + \Cal{B}(V,\phi_0)       \tag {1.82c}
\endalign
$$
(for $a \neq 0$)
is a Poisson map with respect to the Lie Poisson brackets (1.24) and (1.77)
on $(\wt {\grg} \oplus \wt {\grg})^{\wedge *}$ and
diff$_1^{\ \wedge (l_1 l_2 l_3) *}$.
The Hamiltonian flow generated by
$$
\hat J_{(\alpha,\beta,a)}^d  :=  \langle \hat J^d, (\alpha,\beta,a)
\rangle_{\Cal D}
$$
generates the action (1.80a-c) for the 1-parameter subgroup $\text
\{exp[-t (\alpha,\beta,a)]\}$.  Thus,  we have the following commuting
diagram of equivariant moment maps:
$$
\gather
\qquad \qquad T^*\wt{G} \ @>\quad  \hat{J}^{\Cal D} \quad >>
\  \text{diff}_{1}^{\ \wedge (l_1 l_2 l_3) *} \\
{\hat{J}}^{LR} \raise 3pt \hbox{$\searrow$}
\qquad \quad \raise 3pt \hbox{$\nearrow$}
\hat{J}^{d} \; \; \;
\\
\vspace{-2pt}\
(\widetilde{\frak g} \oplus {\widetilde{\frak g}})^{\wedge *}
\endgather
$$
\endproclaim
\demo
{Proof}  The factorization property (1.81) is seen directly from
eqs\. (1.63b,c).  The fact that the map $\hat J^d$ is Poisson with respect to
the Lie
Poisson structures on
$(\wt {\grg} \oplus \wt {\grg})^{\wedge (l_1 l_2 l_3) *}$ and
$\text{diff}_1^{\ \wedge *}$ is proved by the identical computation used
in proving Proposition 1.6.  The vector fields generating the $1-$parameter
subgroups acting through (1.80a-c) are easily computed to be:
$$
\align
X_\alpha^{d 1} =  &- \langle \alpha^{\prime } U + \alpha^{\prime } [\theta_1,
U] + \alpha
U^{\prime } + a \alpha^{\prime \prime } \theta_1, \frac{\delta}{\delta U}
\rangle
\\
&- \langle \alpha^{\prime } V + \alpha^{\prime } [\phi_1 , V] + \alpha
V^{\prime } - a \alpha^{\prime \prime }
\phi_1, \frac {\delta}{ \delta V} \rangle     \tag {1.83a}
\endalign
$$
for the $\Cal {D}_1$-action, and
$$
X_\beta ^{d 0} = - \langle \beta [\theta_0, U] + a \beta^{\prime } \theta_0,
\frac {\delta}{\delta U} \rangle - \langle \beta [\phi_0, V] - a \beta^{\prime
} \phi_0,
\frac {\delta}{\delta V} \rangle      \tag  {1.83b}
$$
for the $\Cal D_0$-action.  Using the definition (1.24) of the Lie Poisson
bracket on $(\wt \grg \oplus \wt \grg)^{\wedge *}$, together with the
Leibnitz rule, we find:
$$
\align
X_\alpha ^{d 1} (\langle \ .\ | (X, Y, b) \rangle) &=
\{\hat J_{(\alpha,0, 0)}^d,
\langle \ .\ | (X,Y,b)\rangle\}  \tag  {1.84a}
\\
X_{\beta }^{d 0} (\langle \ .\ | (X,Y,b)\rangle) &=
\{\hat J_{(0, \beta, 0)}^d,
\langle \ .\  | (X,Y,b)\rangle\}  \ , \tag  {1.84b}
\endalign
$$
and hence $\hat{J}^d$ is the moment map generating the action (1.80a-c)
through Hamiltonian flows.  \hfill $\square$
\enddemo

Note that the moment map  $J^{\Cal D} = (J^0, J^1)$ defined in eqs\. (1.63a-c)
may be expressed in a form that is equally valid whether $k$ vanishes or not:
$$
\align
J^0 = &\Cal{B}(g\mu g^{-1}, \theta_0) + \Cal{B}(-\mu + k g^{-1} g^{\prime },
\phi_0)  \tag  {1.85a}
\\
J^{1} =& - \Cal{B}((g \mu g^{-1})^{\prime }, \theta_1) - \Cal{B}(-\mu^{\prime }
+ k (g^{-1} g^{\prime })^{\prime }, \phi_1)
\\
 &\ + \Cal{B}(g \mu g^{-1}, g^{\prime } g^{-1}) + \frac {k}{2}
\Cal{B}(g^{\prime } g^{-1} , g^{\prime } g^{-1})  \ .
\tag {1.85b}
\endalign
$$

However, for $k=0$, these maps do not factor through
$\hat J^{LR}: T^* \wt G \ra (\wt \grg \oplus \wt \grg)^{\wedge *}$.
In the next subsection, two twisted
$\Cal D_1 \ltimes \Cal D_0$ actions that are Hamiltonian on $T^* \wt G$ will
be considered, whose moment maps, while not factoring through $\hat
J^{LR}$, do instead factor through  the moment map
$\hat J^{LA}: T^* \wt G \ra (\wt \grg \oplus \wt\grg_A)^{\wedge *}$ of
Corollary 1.5.  The first is just a special case of the above, with the
homomorphism $\Phi :  {\Bbb R}^{\times} \times {\Bbb R}^{\times} \ra G$
chosen as trivial, while the second is a new form of ``twisted''
$\Cal D_1 \ltimes \Cal D_0$-action.
\bigskip
\line{1e. \quad{\it  Canonical Symplectic Structure: Factorization Through
$\hat J^{LA}: T^* \wt G \ra (\wt \grg \dotplus \wt \grg_A)^{\wedge *}$ }\hfil}
\medskip
First, consider the $k=0$ case of the $\Cal D_1 \ltimes \Cal D_0$ action
(1.57a-c), with $\Phi$ trivial; i.e. $\phi_0 = \phi_1 = 0$.  From
eq\. (1.85a-b),
we see that the moment map $J^{\Cal D} = (J^0, J^1)$ may be expressed as
$$
\align
J^0 &= \Cal{B}(J^L, \theta_0)  \tag  {1.86a}
\\
J^1 &= - \Cal{B}(J^{L'}, \theta_1) + \Cal{B}(J^L, J^A) \ ,  \tag {1.86b}
\endalign
$$
and thus factors through $\hat J^{LA}: T^* \wt G \ra (\wt \grg \oplus \wt
\grg_A)^{\wedge *}$.  The corresponding induced action on $(\wt \grg
\oplus \wt \grg_A)^{\wedge *}$ is easily computed to be:
$$
(\wt \sigma, f) : (U,V,a) \ra (\check U, \check V,a)   \tag  {1.89a}
$$
where
$$
\align
\check U := &\wt \sigma' \Theta_1 (\wt \sigma') \Theta_0 (f \circ \wt \sigma)
(U \circ \wt \sigma) \Theta_0 (f \circ \wt \sigma)^{-1} \Theta_1
(\wt{\sigma}')^{-1}
 \tag{1.89b}\\
\check V := &\wt \sigma' \Theta_1 (\wt \sigma') \Theta_0 (f \circ \wt \sigma)
(V \circ \wt \sigma) \Theta_0 (f \circ \wt \sigma)^{-1}
\Theta_1 (\wt{\sigma}')^{-1} \\
&\  + \wt \sigma' (\frac {f'}{f} \circ \wt \sigma) \theta_0 + \frac {\wt
\sigma''}{\wt \sigma'} \theta_1
 \ .
\tag {1.89c}
\endalign
$$
The vector field generating the action of the 1-parameter
subgroup  $\{exp[-t(\alpha,\beta)]\}$ is:
$$
X_{(\alpha,\beta)}^{d_a} = X_{\alpha}^{d0_a} + X_{\beta}^{d1_a}  \tag  {1.90a}
$$
where
$$
\align
X_{\beta}^{d0_a} &:= - \langle \beta [\theta_0, V], \frac{\delta}{\delta
V} \rangle - \langle \beta [\theta_0,U] + \beta' \theta_0,
\frac{\delta}{\delta U} \rangle  \tag  {1.90b}
\\
X_{\alpha}^{d1_a} &:= - \langle (\alpha V)' + \alpha' [\theta_1,V], \frac
{\delta}{\delta U} \rangle
- \langle (\alpha U)' + \alpha' [\theta_1, U] + \alpha'' \theta_1, \frac
{\delta}{\delta U} \rangle   \ .  \tag {1.90c}
\endalign
$$
The action (1.89a-c) is again Hamiltonian, and generated by the moment map
$$
\align
J^{d_a} : (\wt \grg \dotplus \wt \grg_A)^{\wedge *} &\lra \text{diff}_1^{\ *}
\\
J^{d_a} : (U,V,a) &\lra (J^{d0_a}, J^{d1_a})  \ ,  \tag {1.91a}
\endalign
$$
where
$$
\align
J^{d0_a} &:=  \frac{1}{a} \Cal{B}(V,\theta_0)   \tag {1.91b}
\\
J^{d1_a} &:= - \frac{1}{a}\Cal{B}(V', \theta_1) + \frac{1}{a}\Cal{B}(U,V) \ .
 \tag{1.91c}
\endalign
$$
In this case, both the maps $J^{\Cal D} : T^* \wt G \ra \text{diff}_1^{\ *}$
and  $J^{d_a}: (\grg \dotplus \wt \grg_A)^{\wedge *} \ra \text{diff}_1^{\ *}$
are equivariant, satisfying:
$$
\align
\{J_{(\alpha_1,\beta_1)}^{\Cal D}, J_{(\alpha_2, \beta_2)}^{\Cal D} \} &=
J_{(\alpha_1 \alpha_2^{\prime } - \alpha_2 \alpha_1^{\prime } , \alpha_1
\beta_2^{\prime } -
\alpha_2 \beta_1^{\prime } )}^{\Cal D}   \tag {1.92}
\\
\{J_{(\alpha_1,\beta_1)}^{d_a}, J_{(\alpha_2, \beta_2)}^{d_a} \} &=
J_{(\alpha_1 \alpha_2^{\prime } - \alpha_2 \alpha_1^{\prime }, \alpha_1
\beta_2^{\prime } -
\alpha_2 \beta_1^{\prime })}^{d_a}  \tag  {1.93}
\endalign
$$
These are therefore Poisson maps with respect to the Lie Poisson structure
on $\text{diff}_1^{\ *}$ without central extension.
Summarizing, we have:
\proclaim
{Proposition 1.10}  For $k = 0$, the moment map $J^{\Cal D}$ factors into:
$$
J^{\Cal D} = J^{d_a} \circ \hat J^{LA}   \ .     \tag  {1.94}
$$
We thus have the commuting diagram of equivariant moment maps:
$$
\gather
T^*\wt{G}  \ @>\quad  J^{\Cal D} \quad >>
\  \text{diff}_{1}^{\ *} \\
{\hat{J}}^{LA} \raise 3pt \hbox{$\searrow$}
\qquad \quad \raise 3pt \hbox{$\nearrow$}
J^{d_a} \; \; \;
\\
\vspace{-2pt}\
(\widetilde{\frak g} \dotplus {\widetilde{\frak g}}_A)^{\wedge *}
\endgather
$$
where $J^{\Cal D}$ generates the $\Cal D_1 \ltimes \Cal D_0$ action (1.57a-c),
with $k=0, \ \phi_0 = \phi_1 = 0$, and $J^{d_A}$ generates the action
(1.89a-c).
\endproclaim
Another ``twisted'' action of $\Cal D_1 \ltimes \Cal D_0$, or rather of the
subgroup $\Cal D_1 \ltimes \Cal D_0^+  \subset \Cal D_1 \ltimes \Cal D_0$
consisting of $\{(\wt \sigma, f), f > 0\}$, may be defined on $T^*\wt G$, one
which is nonequivariant, but still factors through
$\hat J^{LA} : T^* \wt G \ra (\grg \dotplus \wt \grg_A)^{\wedge *}$.
Let  $\Theta=(\Theta_0, \Theta_1) : \Bbb{R}^{\times} \times \Bbb{R}^{\times}
\ra G$  again be a homomorphism, with derivative at  the identity (1, 1)
 identified,  as in eq\.~(1.55), with a pair $\theta_0, \theta_1\in \grg$ of
commuting elements of $\grg$.  Define a right action
$\Cal D_1 \ltimes \Cal D_0^+ : T^*
\wt G \ra T^* \wt G$ by:
$$
(\wt \sigma, f) : (g,\mu) \lra (\Theta_1 (\wt \sigma') (g \circ \wt \sigma)
,\  \wt\sigma' (\mu \circ \wt \sigma) + [\text{ln} (f \circ \wt \sigma)
(g \circ \wt\sigma)^{-1} \Theta_0 (g \circ \wt \sigma)]')  \ . \tag  {1.95}
$$
It is easily verified that this does compose correctly to define a right
action of $\Cal D_1 \ltimes \Cal D_0^+$, and that this action leaves
invariant the canonical symplectic form $\o_0$ on $T^* \wt G$.  The
infinitesimal action obtained by differentiating the flow of
$\{\text{exp}[-t(\alpha,\beta)]\}$ given by (1.95) is represented by the vector
field
$$
X_{(\alpha,\beta)}^{\Cal D_A} := X_{\alpha}^{1_A} + X_{\beta}^{0_A}  \ ,
\tag {1.96a}
$$
where
$$
X_{\beta}^{0_A} := X_{\beta \theta_0}^A   \tag  {1.96b}
$$
generates the $\Cal D_0^+$ action,  and
$$
X_{\alpha}^{1_A} :=  - \langle \alpha g', \frac {\delta}{\delta g} \rangle
- \langle (\alpha \mu)', \frac {\delta}{\delta \mu} \rangle
 + X_{\alpha' \theta_1}^L   \tag  {1.96c}
$$
generates the $\Cal D_1$-action.  Thus
$$
\align
X^{\Cal D_A} : \text{diff}_1 & \lra \chi (T^* \wt G)
\\
X^{\Cal D_A} : (\alpha, \beta) & \lmt X_{(\alpha, \beta)}^{\Cal D_A}  \tag
{1.97}
\endalign
$$
defines a homomorphism to the Lie algebra $\chi (T^* \wt G)$ of functional
derivations (vector fields) on $T^* \wt G$.  The Hamiltonian properties of
this action are given in the following
\proclaim
{Proposition 1.11}  For $k = 0$, the $\Cal D_1 \ltimes \Cal D_0^+$-action
(1.95) is Hamiltonian, and generated by the moment map
$$
\align
J^{\Cal D_A} : T^* \wt G & \lra \text{diff}_1^{\ *}\\
J^{\Cal D_A} : (g,\mu) & \lmt (J^{1_A}, J^{0_A}) \ ,  \tag  {1.98a}
\endalign
$$
where
$$
\align
J^{0_A} & = \Cal{B}(J^A, \theta_0)   \tag  {1.98b}
\\
J^{1_A} & = - \Cal{B}(J^{L'}, \theta_1) + \Cal{B}(J^L, J^A) \ . \tag  {1.98c}
\endalign
$$
This map is nonequivariant, satisfying the Poisson bracket relations:
$$
\{ J_{(\alpha_1, \beta_1)}^{\Cal D_A}, J_{(\alpha_2, \beta_2)}^{\Cal D_A} \}
 =  J_{(\alpha_1 \alpha_2' - \alpha_2\alpha_1', \alpha_1\beta_2' -
\alpha_2\beta_1')}^{\Cal D_A}
 + \Cal{B}(\theta_0, \theta_1) [\langle \alpha_1', \beta_2' \rangle - \langle
\alpha_2',  \beta_1' \rangle ]  \ ,  \tag {1.99}
$$
where
$$
J_{(\alpha, \beta)}^{\Cal D_A} := \langle J^{\Cal D_A}, (\alpha,\beta)
\rangle =
\langle J^{1_A}, \alpha \rangle + \langle J^{0_A}, \beta \rangle   \ .
\tag  {1.100}
$$
\endproclaim
\demo {Proof}  The equalities
$$
\align
X_{\beta}^{0_A}  \bl \o_0 = & - \delta (J_{\beta \theta_0}^A)   \tag {1.101a}
\\
X_{\alpha}^{1_A} \bl \o_0 & = - \delta (J_{\alpha' \theta_1}^L)  -  \delta
\langle \alpha J^L, J^A \rangle  \ ,   \tag  {1.101b}
\endalign
$$
which imply the formulae (1.98b,c) for the moment map $J^{\Cal D_A}$ are
directly verified from the definitions (1.96a,b), (1.8), (1.13a) and (1.30) of
$X^{0_A}, X^{1_A}, \o_0, J^L$ and $J^A$, respectively.  The Poisson brackets
(1.99) are equivalent to the relations:
$$
\align
\{ J_{\alpha_1}^{1_A}, J_{\alpha_2}^{1_A} \}  & = X_{\alpha_1}^{1_A}
(J_{\alpha_2}^{1_A})
\\
& = \{ J_{\alpha_1' \theta_1}^L + \langle \alpha_1 J^L, J^A \rangle,
J_{\alpha_2' \theta_1} + \langle \alpha_2 J^L, J^A \rangle \}
\\
& = J_{\alpha_1 \alpha_2' - \alpha_2\alpha_1'}^{1_A}     \tag  {1.102a}
\\
\{ J_{\alpha}^{1_A} , J_\beta^{0_A} \} & = - X_\beta^{0_A} (J_\alpha^{1_A})
\\
& = \{J_{\alpha' \theta_1}^L + \langle \alpha J^L, J^A \rangle , J_{\beta
\theta_0}^A \}
\\
& = J_{\alpha \beta' \theta_0}^A + \Cal{B}(\theta_0, \theta_1) \langle \alpha'
,
\beta' \rangle    \tag  {1.102b}
\\
\{ J_{\beta_1}^{0_A} , J_{\beta_2}^{0_A} \} & = X_{\beta_1}^{0_A}
(J_{\beta_2}^{0_A})
\\
& = \{ J_{\beta_1 \theta_0}^A, J_{\beta_2 \theta_0}^A \}    \tag  {1.102c}
\\
& = 0 \ ,
\endalign
$$
where
$$
J_{\alpha}^{1_A} := \langle J^{1_A} , \alpha \rangle,\quad
J_{\beta}^{0_A} := \langle J^{0_A} \ , \beta \rangle
  \ .         \tag  {1.103}
$$
These may be verified either directly from the definitions (1.96b,c),
(1.98a,b) of $X_{\beta}^{0_A}$, $X_{\alpha}^{1_A}$, $\ J^{0_A}$ and
$J^{1_A}$, or with the help of the Poisson bracket relations:
$$
\align
\{ J_{\alpha_1^{\prime } \theta_1}^L , \langle \alpha_2 J^L, J^A \rangle \} & =
-
J_{\alpha_1^{\prime \prime } \alpha_2 \theta_1}^L    \tag {1.104a}
\\
\{ \langle \alpha_1 J^L,  J^A \rangle , \langle \alpha_2 J^L, J^A \rangle \} &
= \langle (\alpha_1 \alpha_2^{\prime } - \alpha_2 \alpha_1^{\prime }) J^L, J^A
\rangle \ ,
\tag{1.104b}
\endalign
$$
which follow from eqs\.  (1.40a-c), together with the Leibnitz rule.
\hfill $\square$
\enddemo

Since the cocycle $c_2^{\Cal D}$ (eq\. (1.70b)) enters in (1.99) we proceed, as
usual, defining the central extension $\text{diff}_1^{\ \wedge(0, -l, 0)}$ as
in eq\.  (1.73),
with
$$
l_1 = l_3 = 0  , \qquad    l_2 = \Cal{B}(\theta_1, \theta_0) =:-l \ ,
\tag  {1.105}
$$
which gives the extended coadjoint representation
$$
\hat ad_{(\alpha,\beta,b)}^* (v,w,a)  = (v'\alpha + 2v\alpha' + w \beta' +
al \beta'',\ (w \alpha)' - al \alpha'', 0)    \ .       \tag  {1.106}
$$
We then have
\proclaim
{Corollary 1.12}  The extended moment map
$$
\align
\hat J^{\Cal D_A} : T^* \wt G & \lra \text{diff}_1^{\ \wedge(0, -l, 0) *}
\\
\hat J^{\Cal D_A} : (g, \mu) & \lmt (J^{\Cal D_A}, 1)\ , \tag{1.107}
\endalign
$$
which also generates the $\Cal D_1 \ltimes \Cal D_0^+$-action (1.95), is
equivariant with respect to the extended coadjoint action (1.106).
It is thus a Poisson map with respect to the Lie Poisson structure
determined by the Poisson brackets
$$
\align
\{ \langle \ .\  | (\alpha_1,\beta_1,b_1) \rangle_{\Cal D} ,
\langle \ .\  |
(\alpha_2, \beta_2, b_2) \rangle_{\Cal D}\} |_{(v,w,a)}
=  &\langle v, \alpha_1 \alpha_2' - \alpha_2 \alpha_1' \rangle   + \langle  w,
\alpha_1 \beta_2' - \alpha_2 \beta_1' \rangle
\\
\ & - al ( \langle \alpha_1', \beta_2' \rangle - \langle \alpha_2' , \
\beta_1'
\rangle)   \ . \tag{1.108}
\endalign
$$
\endproclaim
{}From formulae (1.98b,c), we see that  the moment map
$\hat J^{\Cal D_A} : T^* \wt G \ra \text{diff}_1^{\ \wedge(0 l 0) *}$ factors
through
$\hat J^{LA} : T^* \wt G \ra (\wt \grg \dotplus \wt \grg_A)^{\wedge *}$,
suggesting again that the $\Cal D_1 \ltimes \Cal D_0^+$-action (1.95)
induces a Hamiltonian action on $(\wt \grg \dotplus \wt \grg_A)^{\wedge *}$.
This is easily computed, and the result is summarized as follows. (Eqs\.
(1.110b,c) for the case $\th_0 = - \th_1 = \th$ will reappear as eqs\.
(2.42a,b)
in Section 2.)
\proclaim
{Proposition 1.13}  The moment map $\hat J^{LA} : T^* \wt G \ra (\grg
\dotplus \grg_A)^{\wedge *}$ is equivariant with respect to the $\Cal D_1
\ltimes \Cal D_0^+$ action (1.95) on $T^* \wt G$ and the following action on
$(\grg \dotplus \grg_A)^{\wedge *}$:
$$
\align
(\wt \sigma, f) : (\wt \grg \dotplus \wt {\gr{ g}}_A)^{\wedge  *}
& \lra (\wt {\grg} \dotplus \wt \grg_A)^{\wedge *}
\\
(\wt \sigma, f) : (U,V,a) & \lmt (\check U, \check V, a)  \ , \tag  {1.109a}
\endalign
$$
where
$$
\align
\check U & = \wt \sigma \Theta_1 (\wt \sigma') (U \circ \wt \sigma)
 \Theta_1 (\wt \sigma')^{-1}
+ \frac{\wt \sigma''}{\wt \sigma'}  \theta_1
 \tag  {1.109b}
\\
\check V & = \wt \sigma' \Theta_1(\wt \sigma') \left[(V \circ \wt \sigma +
\text{ln} (f \circ \wt \sigma) [ \theta_0, V \circ \wt \sigma]\right]
 \Theta_1 (\wt \sigma')^{-1}
+ \frac{(f \circ \wt \sigma)'}{f \circ \wt \sigma}  \theta_0
 \ . \tag  {1.109c}
\endalign
$$
This action is Hamiltonian, and generated by the equivariant moment map
$$
\align
\hat J^{d_A} : (\wt \grg \dotplus \wt \grg_A)^{\wedge *} & \lra
\text{diff}_1^{\ \wedge(0,-l,0) *}
\\
\hat J^{d_A} : (U,V,a) & \lmt (J^{d1_A}, J^{d0_A}, 1)  \ ,   \tag  {1.110a}
\endalign
$$
where
$$
\align
J^{d0_A} &:= \Cal{B}(U,\theta_0)   \tag  {1.110b}
\\
J^{\alpha 1_A} &:= \frac{1}{a} [ - \Cal{B}(V',\theta_1)
+ \Cal{B}(U,V)]  \ .   \tag  {1.110c}
\endalign
$$
This is therefore a Poisson map with respect to the Lie Poisson structure
determined by the Poisson brackets (1.108).
\endproclaim
\demo
{Proof}  Differentiating along the flow generated by the action (1.109a-c)
corresponding to the 1-parameter group $\{\text{exp}[-t(\alpha, \beta)]\}$
gives the representation
$$
\align
X^{d_A} :\text{diff}_1 & \lra \chi ((\wt \grg \dotplus \wt \grg_A)^{\wedge *})
\\
X^{d_A} : (\alpha, \beta) & \lmt X_{(\alpha, \beta)}^{d_A} :=
X_{\alpha}^{d1_A} + X_{\beta}^{d0_A}   \tag  {1.111a}
\endalign
$$
of $\text{diff}_1$ by functional derivations (vector fields) on
$(\grg \dotplus \grg_A)^{\wedge *}$ defined by:
$$
\align
X_{\beta}^{d0_A} & := - \langle (\beta [\theta_0,U] + \beta' \theta_0)
,\frac{\delta}{\delta V} \rangle   \tag  {1.111b}
\\
X_{\alpha}^{d1_A} & := - \langle (\alpha V)' + \alpha' [\theta_1, V] ,
\frac{\delta}{\delta V} \rangle
- \langle (\alpha U)' + \alpha' [\theta_1, U] + \alpha'' \theta_1,
\frac{\delta}{\delta U} \rangle   \ .  \tag  {1.111c}
\endalign
$$
By explicit evaluation, and use of the Poisson bracket relations (1.40a-c) and
the Leibnitz rule, we find
$$
\align
\{ J_{\beta}^{d0_A}, \langle  \ .\ | (X, Y,b)\rangle \} & = X_{\beta}^{d0_A}
(\langle \ .\  | (X,Y,b) \rangle)  \tag  {1.112a}
\\
\{ J_{\alpha}^{d1_A}, \langle \ .\  | (X,Y,b) \rangle \} & =
X_{\alpha}^{d1_A} (\langle  \ .\ | (X,Y,b) \rangle )  \tag  {1.112b}
\\
\{ J_{\beta_1}^{d0_A}, J_{\beta_2}^{d0_A} \} & = X_{\beta_1}^{d0_A}
(J_{\beta_2}^{d0_A}) = 0    \tag   {1.112c }
\\
\{ J_{\alpha}^{d1_A}, J_{\beta}^{d0_A} \} & = - X_{\beta}^{d0_A}
(J_{\alpha}^{d1_A})  = J_{\alpha \beta'}^{d0_A} + \Cal{B}(\theta_0, \theta_1)
\langle \alpha', \beta' \rangle
 \tag  {1.112d}
\\
\{ J_{\alpha_1}^{d1_A} , J_{\alpha_2}^{d1_A} \} & = X_{\alpha_1}^{d1_A}
(J_{\alpha_2}^{d1_A}) = J_{(\alpha_1 \alpha_2' - \alpha_2 \alpha_1')}^{d1_A}
\ ,   \tag  {1.112e}
\endalign
$$
showing both that $\hat J^{d_A} = (J^{d1_A}, J^{d0_A} , 1)$ is the moment
map generating the action (1.109a-c), and that it is equivariant, hence
preserving the respective Lie Poisson structures on $(\wt \grg \dotplus \wt
\grg_A)^{\wedge *}$ and $\text{diff}_1^{\ \wedge(0,-l,0) *}$ defined by eqs\.
(1.39)
and (1.108).
\hfill   $\square$
\enddemo

For the applications to integrable systems to be discussed in Section 2,
a particular case of the  $\Cal D_1\ltimes \Cal D_0^+$ action (1.109a-c) and
moment map (1.110a-c) will be used; namely, when
$$
\theta_0 = - \theta_1 =: \theta \ ,   \tag{1.113a}
$$
and hence
$$
\Cal B (\theta, \theta) = l   \ . \tag{1.113b}
$$

Finally, combining the results of Propositions 1.11 and 1.13 and Corollary
1.12, we get:
\proclaim
{Theorem 1.14}  For $k = 0$, the moment map $\hat J^{\Cal D_A} : T^* \wt G
\ra \text{diff}_1^{\ \wedge (0, -l, 0)*}$ factors into
$$
\hat J^{\Cal D_A} = \hat J^{d_A} \circ \hat J^{LA} ,      \tag    {1.114}
$$
giving the following commuting diagram of equivariant moment maps:
$$
\gather
T^*\wt G  \ @>\quad  \hat{J}^{\Cal D_A} \quad >>
\  \text{diff}_{1}^{\ \wedge(0,-l,0) *} \\
{\hat{J}}^{LA} \raise 3pt \hbox{$\searrow$}
\qquad \qquad \raise 3pt \hbox{$\nearrow$}
{\hat{J}}^{d} \qquad \qquad \\
\vspace{-2pt}\
(\widetilde{\frak g} \dotplus {\widetilde{\frak g}}_A)^{\wedge *}
\qquad \quad
\endgather
$$
\endproclaim
\bigskip
\bigskip
\newcount\refn
\refn=0
\def\ref{\vskip 18 pt \global\advance\refn by1\item 
{{\the\refn.}}}
\newcount\notenumber
\def\clearnotenumber{\notenumber=0}
\def\note{\advance\notenumber by 1 \footnote{$^{\the\notenumber}$}}
\clearnotenumber

\def \nd{\noindent}
\def\leaderfill{\leaders \hbox to 10 pt{\hss.\hss}\hfill}

\def\pmb#1{\setbox0=\hbox{#1}%
  \kern-.025em\copy0\kern-\wd0
  \kern.05em\copy0\kern-\wd0
  \kern-.025em\raise.0433em\box0 }
\def\i{\item}

\def\pvo{\par \vskip 15 pt}

\def\boxit#1{\vbox{\hrule\hbox{\vrule\kern3pt
      \vbox{\kern3pt#1\kern3pt}\kern3pt\vrule}\hrule}}
\def\Z{{\hbox{{\bf Z}}}}

\def\N{{\hbox{{\bf N}}}}

\def\cX{{\cal X}}

\def\i{\item}

\def\dps{\displaystyle}

\def\pvo{\par \vskip 15 pt}

\def\boxit#1{\vbox{\hrule\hbox{\vrule\kern3pt
      \vbox{\kern3pt#1\kern3pt}\kern3pt\vrule}\hrule}}

\newcount\chapno
\chapno=0
\def\ch{\global\advance\chapno by 1{\phantom {\the\chapno}}}
\newcount\exno
\exno=0
\def\ex{\global\advance\exno by1{(\the\exno)}}
\newcount\refn
\refn=0
\def\ref{\bigskip\global\advance\refn by1\item
{{\the\refn.}}}

\def \nd{\noindent}

\def \i{\item}

\def \1{\hbox to .75truein { }}
\def \2{\hbox to 2.5truein{ }}
\def \3{\hbox to 6.5truein{ }}
\def \4{\hbox to 2.0truein{ }}
\def \5{\hbox to 1.0truein{ }}
\def \6{\hbox to .25truein{ }}
\def \7{\hbox to 3.0truein{ }}

\def\leaderfill{\leaders \hbox to 10 pt{\hss.\hss}\hfill}

\def\pmb#1{\setbox0=\hbox{#1}%
\kern-.025em\copy0\kern-\wd0
\kern.05em\copy0\kern-\wd0
\kern-.025em\raise.0433em\box0 }

\def\dps{\displaystyle}


\def\vline{\vrule height 17pt depth 5pt}
\def\back{\noalign{\vskip-3pt}}
\def\bback{\noalign{\vskip-11pt}}
\def\3{\noalign {\vskip 3pt}}
\def\2{\hskip .25truein}
\def\5{\hskip .50truein}

\def\lim{{\rom lim}}

\def\wG{\widetilde{G}}

\def\N{{\bold {N}}}

\def\Z{{\bold {Z}}}

\def\V{{\bold {V}}}
\def\U{{\bold {U}}}
\def\b1{{\bold {1}}}
\def\o{{\bold {0}}}
\def\pmtheta{\pmb{$\theta$}}
\def\pmPhi{\pmb{$\Phi$}}
\def\pmphi{\pmb{$\phi$}}

\def\pmg{\pmb{$\gamma$}}

\def\pmq{\pmb{$q$}}

\def\pmu{\pmb{$u$}}

\def\B{{\Cal {B}}}

\def\G{{\Cal {G}}}
\def\H{{\Cal {H}}}

\def\cX{{\Cal {X}}}

\def\const{{\rom const}}

\def \Xb{\bar{X}}

\line{{\bf 2. \quad Integrable Systems }\hfil}
\bigskip
\line{{\it 2a. \quad  Algebraic Language}\hfil}
\medskip
The preceding section emphasized the symplectic geometry of
$T^*\wG$ and Lie Poisson structures.  In modern theories of integrable
systems in $1+1$ dimensions, one often deals with local evolution
equations, for which the language of differential algebra is the most
efficient.  On a practical level, the translation from functional to
algebraic language consists of replacing:
\smallskip \nd
\i{1)}Functionals by their densities, and integration by parts by
equivalence modulo ``divergences'' for these densities.  (The algebraic
calculus of variation results thereby.)
\smallskip
\nd
\i{2)}Symplectic forms and Poisson brackets by a Hamiltonian map $H
\mapsto X_H$ assigning to any Hamiltonian density  $H$ the
evolution derivation $X_H$ via the rule
$$
(\pmq_{,t} = ) \ ~X_H (\pmq) = B\left({\delta H \over \delta \pmq}\right)
 \ , \tag{2.1}
$$
where $\pmq$ is a column--vector of the basic variables (``fields'')
chosen in a fixed basis, with $i^{\text{th}}$ component $q_i$;
 ${\delta H \over \delta \pmq}$ is the column--vector of variational
derivatives of $H$, and $B$ is a skew symmetric matrix differential operator
whose properties guarantee the Jacobi identity.

In the loop space setting,  the most frequently met structure consists of
Lie Poisson brackets associated with centrally extended Lie algebras.
 This is translated into the following algebraic construction.  Denote
by $K$ a differential algebra with derivation $\partial$. (In Section
1, these were $K=C^\infty (\Bbb R)$ or $C^\infty (S^1)$ and $\partial =
\partial/\partial \sigma$.)  Let $\G = K^{\overline{n}} $,
$\overline{n} \in \N \cup \{ \infty \}$, be a differential Lie algebra
consisting of
column vectors of dimension $\overline{n}$ with entries in $K$.  The
commutator in $\G$ is of the form:
$$
\align
[X,Y]_k &=
\sum_{ijpr}c^k_{ijpr}\partial^p(X_i)\partial^r(Y_j) \qquad
(\text{finite sum } \forall \ k) \ , \tag{2.2} \\
  X, Y & \in \G   \ ,
\endalign
$$
where $c^k_{ijpr} \in K$ are structure elements defining $\G$ and
$X_i \in K$ denotes the $i^{th}$ component of $X \in K^{\overline{n}}$.  Let
$\Omega$
be a (generalized) $2$--cocycle on $\G$.  This means that $\Omega: \G
\times \G \rightarrow K$ is a bilinear skewsymmetric differential
operator satisfying
$$
\align
\Omega([X,Y],Z) +  \Omega ([Y,Z], X) & + \Omega ([Z,X],Y) \sim 0
\tag{2.3a} \\
\Omega(X,Y) & \sim - \Omega (Y,X)  \ ,  \tag{2.3b} \\
 \forall \  X, Y, Z  & \in \G \ ,
\endalign
$$
where $(\cdot) \sim 0$ means that $(\cdot) \in \text{Im} \partial$.
In the loop algebra setting, one associates with this data the centrally
extended algebra $\G^{\wedge} = \G \oplus \Bbb R$, with Lie bracket
$$
[(X,a), (Y,b)] = ([X,Y], \int \Omega (X,Y) d \sigma), \ \ \ X,
Y \in \G, \ \ a, b \in \Bbb R. \tag{2.4}
$$
In the algebraic approach, one associates to the pair $(\G, \Omega)$ an
{\it affine Hamiltonian matrix} (of differential operators)
$$
B = B (\G) + b_\Omega ,  \tag{2.5}
$$
where the linear  part $B(\G)$  (in the basic variables $\pmq$)
of the Hamiltonian matrix $B$ is defined by the relation
$$
[B(\G) (X)]^t Y \sim \pmq^t [X,Y]: = \sum_{k = 1}^{\overline{n}} q_k
[X,Y]_k, \ \ \ \ \forall \ X, Y \in \G \ . \tag{2.6a}
$$
 A $\pmq$--independent matrix differential operator $b_\Omega$ is associated to
the
bilinear form $\Omega$ via the rule
$$
[b_\Omega (X)]^t Y \sim \Omega (X,Y), \qquad \forall  \ X, Y \in \G \ .
\tag{2.6b}
$$

One of the basic features of the algebraic Hamiltonian formalism is that
there is a natural one-to-one correspondence, given by the formulae
(2.6a,b), between  pairs $(\G, \Omega)$ and affine Hamiltonian
matrices.  (The latter encompass more general situations
than the geometric ones, where the Hamiltonian operators are of order
zero in $\partial$. For example, Lie algebras of differential and
pseudodifferential operators have no natural local description as the
infinitesimal form of groups.) The geometric Lie Poisson brackets of
the previous section, which express the multiplication rule (2.4)
in the language of linear functionals can be extracted from formula (2.5)
via the following computation.
Let $X, Y \in \G$, and let $H= \pmq^t X, \  F  = \pmq^t Y$
be linear Hamiltonians.  Then
$$
\align
\{\pmq^t X ,  \pmq^t Y \}& = \{H,F\} =: X_H (F) \sim {\delta F
\over \delta \pmq^t} X_H (\pmq) = {\delta F \over \delta \pmq^t}
B \left({\delta H \over \delta \pmq}\right) \\
&= Y^t [B(\G) + b_\Omega] (X)  \sim \pmq^t [X,Y]
+ \Omega (X,Y) \ . \tag{2.7}
\endalign
$$
{\it Remark 2.1}. \ The recipe (2.5), (2.6a,b) handles
situations, like that of formula (2.4), where the centrally extended
Lie algebra in question has components of different differential
dimension; $X$ and $Y$ are functions, $a$ and $b$ are numbers.  When
$\G$ is a Lie algebra over $\Bbb R$ (or $\Bbb{C}$, etc.) rather than over $K$,
and $\Omega$ is a true $2-$cocycle, not a generalized one (i.e. one
has  an equality sign instead of  the $\sim$ sign in formulae
(2.2), (2.3)), then $\G^{\wedge} \sim \G \oplus \Bbb R$ is again a Lie algebra
over $\Bbb R$.  Treating it as a new Lie algebra, we get from formulae
(2.6a,b) that
$$
B (\G^{\wedge}) = \bordermatrix{&\pmq&&a\cr
\pmq&B(\G)+ ab_\Omega& \vline & \o\cr
\bback
&\multispan3\hrulefill\cr
\back
a &\o^t& \vline &0\cr}, \tag {2.8})
$$
where $a$ is the extra coordinate on the $\Bbb R$-part of
$\G^{\wedge \ast} \sim\G^{\ast} \oplus \Bbb R$.  Thus, we can let
$$
a = \const.  \tag {2.9}
$$
in formula $(2.8)$ and what remains from $B(\G^{\wedge})$ is our
universal formula $(2.5)$ (for $a=1$).

Let us consider, as an illustration, the case of the Lie algebra
$\widetilde{\grg}$ (or $L \grg$) of the previous section.  Fix a basis in
$\grg$.  Then $\widetilde{\grg} =
K^{\overline{n}}$, where $\overline{n} = \text{dim} \grg$.  Let the
Ad--invariant form $\B$ on $\grg$ be given, in the chosen basis, by a
symmetric matrix $\overline{\B}$:
$$\B(x,y) = y^t \overline{\B} x, \ \ \ \ x, y \in \grg, \tag {2.10}
$$
and let $c_{ij}^k$ be the structure constants of $\grg$ in the same
basis:
$$[x,y]_k = \sum_{jk} c_{ij}^k x_j y_i. \tag {2.11}$$
Then, for both $\grg$ (over $\Bbb R$) and $\widetilde{\grg}$ (over $K$)
$$ B(\grg)_{ij}=
B(\wt{\grg})_{ij} = \sum_{kj} c_{ij}^k q_k. \tag {2.12}
$$
Also, since
$$\Omega(X,Y) = \B (\partial(X), Y) = Y^t \overline{\B} \partial (X),
\tag {2.13}$$
we see that
$$b_\Omega = \overline{\B} \partial, \tag {2.14}$$
so that, finally,
$$B_{ij} = \sum c_{ij}^k q_k + \overline{\B}_{ij} \partial. \tag{2.15}
$$
If an orthonormal basis for $\grg$  is chosen (with
respect to $\B$), we have
$$\overline{\B}_{ij} = \delta_{ij} \ \ , \ \ \ c_{ij}^k = c_{jk}^i \ ,
\tag{2.16}
$$
so that the equations  of  motion for a  Hamiltonian $H$ are
$$
\dot{q}_i = \sum_j B_{ij} \left({\delta H \over \delta
q_j}\right) = \sum _{jk} c_{ij}^k q_k {\delta H \over \delta q_j} +
\left({\delta H \over \delta q_i}\right)^\prime = - \left[\pmq, {\delta H
\over \delta \pmq}\right]_i + \left[\left({\delta H \over \delta
\pmq}\right)^\prime \right]_i \ , \tag{2.17a}
$$
i.e.
$$
\dot{\pmq} = (- [ \pmq, \ \ ] + \partial \b1) \left({\delta H \over
\delta \pmq}\right). \tag{2.17b}
$$
In other words,
$$
B = - [ \pmq, \ \ ] + \partial \b1. \tag{2.18}
$$
If the chosen basis of $\grg$ is not orthonormal, then one has the
familiar form
$$
B(\ \cdot \ )= - ad^\ast_{( \ \cdot \ )} \pmq+ \overline{\B} \partial\ .
\tag{2.19}
$$

Finally, we discuss Hamiltonian maps.  Suppose $B$ is a Hamiltonian
matrix over the ring $C_q = K[\pmq, \pmq^\prime, \ldots]$, and let
$B_1$ be a Hamiltonian matrix over another ring $C_u$. A
homomorphism of differential rings (over $K$)
$$\Phi: C_u \rightarrow C_q \tag{2.20}
$$
which commutes with $\partial$ is called a Hamiltonian map if
$$\Phi X_H = X_{\Phi (H)} \Phi \ ,  \qquad \forall  \ H \in C_u.
 \tag{2.21}
$$
(This map should be viewed as dual to a Poisson map on the  fields $\pmq
\mapsto \pmu$ which was the object of study throughout Section 1.) In terms of
the
Hamiltonian matrices $B$ and
$B_1$, the compatibility condition (2.21) is expressed by the equality
$$\
\Phi(B_1) = D(\pmPhi) B D (\pmPhi)^\dagger, \tag{2.22}
$$
where:
$$\pmPhi: = \Phi (\pmu), \ \ \Phi_\alpha = \Phi(u_{\alpha}) \ , \eqno
(2.23)
$$
$D(\pmPhi)$ is the Fr$\acute{\rom{e}}$chet derivative of $\pmPhi$:
$$D(\pmPhi)_{\alpha i} = \sum_\ell {\partial \Phi_\alpha \over
\partial q_i^{(\ell)}} \partial^\ell \ , \tag{2.24}
$$
and `${}^\dagger$'  denotes adjoint.

As an example,
$$
2B_1 = 2 (u \partial + \partial u) - l \partial^3 \tag{2.25}
$$
defines the differential algebraic version of the Virasoro algebra,
denoted here $C_u$ (equiv.  $(\text{diff } S^1)^{\wedge}$)
associated with the generalized $2-$cocycle
$$
\Omega(\alpha_1,\alpha_2) = {l \over 2} \alpha_1 \partial^3(\alpha_2)
\tag{2.26}
$$
on the Lie algebra of vector fields on $ \Bbb R$ (or $S^1$).

Consider the ring homomorphism generated by
$$
\Phi(u) = \pmtheta^t \pmq^\prime + \pmq^t \pmq \tag{2.27}
$$
from $C_u$ into $C_q$, where $\pmtheta$ is a fixed (constant) vector
in $\grg$, with the Hamiltonian matrix $B$ in the ring $C_q$ given by
formula $(2.18)$. Then $\Phi$ is a Hamiltonian map, since
$$
\align
D(\pmPhi)BD(\pmPhi)^\dagger & = (\pmtheta^t \partial + 2 \pmq^t)
([- \pmq, \ \ ] + \partial \b1) (-\pmtheta \partial + 2 \pmq) \\
& = (- \partial \pmtheta^t [\pmq, \ \ ] + \pmtheta^t \partial^2 + 2
\pmq^t \partial) ( - \pmtheta \partial + 2 \pmq) \\
& = - \pmtheta^t \pmtheta \partial^3 + 2 \partial^2 \pmtheta^t \pmq - 2
\pmq^t \pmtheta \partial^2 + 4 \pmq^t \partial \pmq\\
&= - \pmtheta^t \pmtheta \partial^3 + 2 (\pmtheta^t \pmq^\prime
\partial + \partial \pmtheta^t \pmq^\prime) + 2 (\pmq^t \pmq \partial
+ \partial \pmq^t \pmq) \\
&= \Phi (- \pmtheta^t \pmtheta \partial^3 + 2u \partial + 2 \partial u)
= \Phi (2B_1) \ ,  \tag{2.28}
\endalign
$$
provided
$$
l = \pmtheta^t \pmtheta \ . \tag{2.29}
$$
Thus, the criterion $(2.22)$ is satisfied. (The geometrical significance
of the Hamiltonian map (2.27) in the loop group setting is explained in
{\bf HK1]}.)

\noindent
{\it Remark 2.2.}\quad To make contact with the notation of the previous
section, we adopt the convention that boldface letters denote column
vectors representing the components of elements either of $\grg$ relative
to some basis, or of $\grg^*$ relative to the dual basis, or elements of the
corresponding loop algebra $\wt{\grg}$ or its dual $\wt{\grg}^*$. Thus,
the elements $\th,\ \phi \in \grg$ are replaced by the column vectors
$\pmtheta, \ \pmphi \in {\Bbb R}^{\text{dim}\grg}$, while the elements $U,\ V
\in \wt{\grg}^*$ are
replaced by the column vectors $\U, \V \in {K}^{\text{dim}\grg}$.
\bigskip
 \noindent
\line{{\it 2b. \quad  Dispersive Wave Systems}\hfil}
\medskip
In Section 1 we derived two families of Poisson  maps:
$$
\hat{J}^d: (\widetilde{\grg} \oplus \widetilde{\grg})^{\wedge \ast}\lra
\text{diff}_1^{\ \wedge(l_1l_2l_3)*}  \ , \tag{2.30}
$$
 defined by eqs\. (1.82a-c) and
$$
\hat{J}^{d_A}: (\widetilde{\grg} \dotplus \widetilde{\grg}_{A})^{\wedge \ast}
\lra \text{diff}_{1}^{\ \wedge(0,-l,0)*} \ , \tag{2.31}
$$
defined by eqs\. (1.91a-c).
In the following, we shall consider two special cases of these maps; for the
 map (2.30), we take the values for $\theta_0, \theta_1, \phi_0, \phi_1$
in eqs\. (1.82a-c) as given by eqs\. (1.78a,b), for which
$(l_1l_2l_3)=(4l, -2l, 0)$, and denote the  corresponding centrally extended
algebra  $\text{diff}_{1(1)}^{\ \wedge} :=\text{diff}_1^{\ \wedge(4l, -2l, 0)}$
for brevity. For the map (2.31), we  choose the values for
 $\theta_0, \theta_1$  in eq\. (1.91a-c) as given by eqs\. (1.113a) and
denote the centrally extended algebra $\text{diff}_{1(0)}^{\ \wedge}
:=\text{diff}_1^{\ \wedge(0,-l,0)}$. Thus, the subscript $(\a) =(1),\ (0)$
distinguishes two different central extensions of the Lie algebra
$\text{diff}_1$. We first convert these maps, according to the lexicon of the
preceding subsection, into algebraic Hamiltonian maps and then derive
integrable hierarchies on
$(\widetilde{\grg} \oplus \widetilde{\grg})^{\wedge \ast}$ and
$(\widetilde{\grg} \dotplus \widetilde{\grg}_{A})^{\wedge \ast}$ from known
integrable hierarchies on $\text{diff}_{1 (\alpha)}^{\ \wedge *}$.

When $(\alpha) =(1)$, we have the Lie algebra $\widetilde{\grg} \oplus
\widetilde{\grg}$ where, choosing normalizations  to correspond to the
fluid dynamical conventions in {\bf[Ku1]}, the commutator
 is
$$
\left[\pmatrix
\Xb_1  \\  {} \\ Y_1
\endpmatrix_,
\pmatrix
\Xb_2 \\ {} \\
Y_2
\endpmatrix \right] =
\pmatrix 2 [\Xb_1, \Xb_2] \\  {}\\ [Y_1, Y_2]
\endpmatrix,  \qquad
\Xb_i, Y_i \in \widetilde{\grg} \ ,  \tag{2.32a}
$$
and the $2-$cocycle is
$$
\Omega(1,2):= \Omega \left(\pmatrix \Xb_1 \cr \cr Y_1\cr\endpmatrix,
 \pmatrix  \Xb_2
\cr \cr Y_2\cr \endpmatrix\right) = 2\Xb^t_1 \Xb^\prime_2 -
{1 \over 2} Y_1^t Y_2^\prime
\tag{2.32b}
$$
Thus, we are working with the  centrally extended algebra
 $(\wt{\grg}\oplus \wt{\grg})^{\wedge}$, with Lie bracket (1.20),
under the identifications $({1\over 2} X, Y) = (\Xb, Y)$ and the $2$-cocycle
$\Omega$ of eq\. (2.32b) is normalized to half $c^{LR}$ of eq\. (1.19).
To simplify calculations, we choose from now on an orthonormal
basis in $\grg$.  Denoting the generic basic variable $\pmq$ by $\U$ and
$\overline{\V}$ (corresponding to $2U$ and $V$ in (1.21)), the Hamiltonian
matrix (2.5) encoding the data (2.32a,b)
is
$$
 B_{(1)} = \pmatrix -2 [ \U, \ \ ] + 2 \partial \b1&&\o\cr
\cr \o&&-[\overline{\V}, \ \ ] - {1 \over 2} \partial \b1 \endpmatrix
\tag{2.33}
$$
This Hamiltonian matrix is thus the algebraic version of the Lie Poisson
bracket (1.24) at $a= {1 \over 2}$.

For the Lie algebra $\text{diff}_1$, we have the commutator
$$
\left[\pmatrix \beta_1 \cr \cr \a_1\endpmatrix _, \pmatrix \beta_2 \cr \cr
\a_2\endpmatrix \right] = \pmatrix \a_1 \beta_2^\prime - \a_2 \beta_1^\prime
\cr \cr \a_1
\a_2^\prime - \a_1^\prime \a_2\endpmatrix , \ \ \ \ \beta_i, \a_i \in K \ ,
\tag{2.34}
$$
and the $2$-cocycle
$$\
\Omega_{(1)} (1,2) = l \left(2\beta_1 \beta_2^\prime + \a_1
\beta_2^{\prime \prime} - \beta_1 \a^{\prime \prime}_2 \right) \tag{2.35}
$$
corresponding to the central extension $\text{diff}_{1(1)}^{\ \wedge}$
(cf\.  eq\. (1.78c)).
Denoting the generic basic variable $\pmq$ as  $h:=v$ and  $u:= w$
 (the standard  notation from fluid dynamics), the corresponding Hamiltonian
matrix is
$$
\overline{B}_{(1)} = \pmatrix 2 l \partial
&& \partial u - l \partial^2 \cr \cr u \partial +
l \partial^2&&h \partial + \partial h
\endpmatrix  \ .
\tag{2.36}
$$
This gives the algebraic counterpart of the Lie Poisson bracket (1.79)
(at $a={1 \over 2}$).

For $\text{diff}_{1(0)}^{\ \wedge}$, we have another
$2$-cocycle  on the Lie algebra $\text{diff}_1$:
$$
\Omega_{(0)} (1,2) = l \left(\a_1 \beta_2^{\prime \prime} -
\beta_1 \a^{\prime \prime}_2 \right) \ . \tag{2.37}
$$
The corresponding Hamiltonian matrix on $\text{diff}_{1(0)}^{\ \wedge \ast}$
is
$$\
 \overline{B}_{(0)} = \pmatrix 0&&\partial u - l \partial^2 \cr \cr u \partial
+ l \partial^2&&h \partial + \partial h \endpmatrix \ .
\tag{2.38}
$$
which is the counterpart of the Lie Poisson bracket (1.108) at $a=1$.

For the Lie algebra $\widetilde{\grg} \dotplus  \widetilde{\grg}_{A}$, we have
the commutator (cf\.  (1.26))
$$\
\left[ \pmatrix Y_1 \cr \cr X_1\endpmatrix _, \pmatrix Y_2 \cr \cr
X_2\endpmatrix  \right] = \pmatrix [X_1, Y_2] - [X_2, Y_1] \cr \cr [X_1, X_2]
\endpmatrix , \ \ \ \ X_i, Y_i \in \widetilde{\grg} \ ,
\tag{2.39}
$$
and the $2-$cocycle (cf\. (1.37))
$$
\Omega(1,2) = Y_1^t X_2^\prime + X_1^t Y^\prime_2 \ ,
\tag{2.40}
$$
so that the corresponding Hamiltonian matrix, in the variables $\pmq =
(\U, \V)$ (which correspond in this case to the pair $(U,V)$ in (1.27)), is
$$
B_{(0)} = \pmatrix \o&&- [\U, \ \ ] + \partial \b1 \cr \cr
-[\U, \ \ ] + \partial \b1&&-[\V, \ \ ]\endpmatrix \ ,
\tag{2.41}
$$
which is the algebraic counterpart of the Lie Poisson bracket (1.47)
at $a=1$.

{}From Theorem 1.9 of the preceding section, with $\phi_0 = 0$,
$\theta_0 = -2\phi_1 =- 2\theta_1 = 2\theta$, $a={1 \over 2}$, and $\theta$
replaced by the corresponding column vector $\pmtheta$, we have
\proclaim
{Proposition 2.1}  The map $\Phi_{(1)}: C_{u,h}
\rightarrow C_{\U, \overline{\V}}$, given on the generators $u$ and
$h$ by the formulae
$$
\align
\Phi_{(1)} (u) &= \pmtheta^t \U     \tag{2.42a}\\
\Phi_{(1)} (h) &= \pmtheta^t ({1 \over 2} \U^\prime +
\overline{\V})^\prime + {1 \over 4} \U^t \U - \overline{\V}^t
\overline{\V} \ ,
\tag{2.42b}
\endalign
$$
is a Hamiltonian map with respect to the Hamiltonian matrices
$\overline{B}_{(1)}$  (2.36) and $B_{(1)}$  (2.33), with
$l:=\pmtheta^t \pmtheta$.
\endproclaim
{}From Proposition 1.13 of the preceding section, with
 $\theta=\theta_0=-\theta_1$, we have
\proclaim {Proposition 2.2}
The map $\Phi_{(0)}: C_{u,h} \rightarrow C_{\U, \V}$, defined by
the formulae
$$
\align
\Phi_{(0)} (u) &= \pmtheta^t \U    \tag{2.43a} \\
\Phi_{(0)} (h) &= \pmtheta^t \V^\prime + \U^t \V  \ , \tag{2.43b}
\endalign
$$
is a Hamiltonian map with respect to  the Hamiltonian matrices
$\overline{B}_{(0)}$  (2.37) and $B_{(0)}$ (2.41), again with
$l:=\pmtheta^t \pmtheta$ .
\endproclaim
These results may also be  verified  directly, of course, by a computation
similar to (2.28).
Until the end of this section, $\pmtheta$ will be taken to have
unit length:
$$
l= \pmtheta^t \pmtheta = 1 \ . \tag{2.44}
$$
In the Hamiltonian structure $\overline{B}_{(1)} \ (2.36)$, the
following sequence of Hamiltonians
$$
\align
H_1 &= h, \qquad  H_2 = uh, \quad \dots   \tag{2.45a}\\
H_n  &:= {1 \over n} \ {\rom{Res}} \ ~(\partial + u + h
\partial^{-1})^n \ , \qquad n \in \N \ ,
\tag{2.45b}
\endalign
$$
where
$$
 {\rom{Res}}  \left(\sum a_i \partial^i \right) : = a_{-1} \ , \tag{2.46}
$$
is known to form a commuting family {\bf [Ku1]}.  The first flow, with $H
= h$, produces the $\sigma$-shift (i.e. $ \dot{\pmq} = \pmq^\prime$,
whatever $\pmq $ is), while the flow of the Hamiltonian
$$
H = {1 \over 2} H_2 = {1 \over 2} uh  \tag{2.47}
$$
produces the equations of dispersive water waves (DWW)
$$
\align
\dot{u}&= ({1 \over 2} u^2 + h - {1 \over 2}
u^\prime)^\prime   \tag{2.48a} \\
\noalign{\vskip 5 pt}
\dot{h}&= (uh + {1 \over 2} h^\prime)^\prime \ .
\tag{2.48b}
\endalign
$$
When $\grg$ is one--dimensional, the Hamiltonian map
$\Phi_{(1)}$ produces an infinite commuting hierarchy of Hamiltonians
$\Phi_{(1)} (H_n)$, called the modified dispersive water wave (mDWW)
hierarchy (in the variables $U, \overline{V} + {1 \over 2} U)$.  For $\grg$
arbitrary,
we again obtain an infinite commuting hierarchy, this time with
respect to the Hamiltonian structure $(2.33)$.  The first flow, with
Hamiltonian
$$\Phi_{(1)} (h) \sim {1 \over 4} \U^t \U - \overline{\V}^t
\overline{\V} \ ,  \tag{2.49}
$$
is just the $\sigma$-shift, as was to be expected.

 Let us compute the equations  of motion for the next flow, which we denote
$\grg$-mDWW  ($\grg$-modified dispersive water waves).
We have, for any $H \in C_{u,h}$:
$$
\pmatrix \dps{{\delta}\strut \over \strut \dps{\delta \U}}
\cr \cr \dps{{\delta} \strut \over \strut \dps{\delta
\overline{\V}}}\endpmatrix \!\! \left(\Phi_{(1)} (H)\right) = D
(\pmPhi_{(1)})^\dagger \Phi_{(1)}\!\! \pmatrix \dps{{\delta} \strut \over
\strut \dps{\delta u}} \cr \cr \dps{{\delta} \strut \over \strut
\dps{\delta h}}\endpmatrix \!\! (H)\ , \tag{2.50}
$$
where
$$\pmPhi_{(1)} = \pmatrix \Phi_{(1)} (u) \cr \cr \Phi_{(1)} (h)\endpmatrix  =
\pmatrix \pmtheta^t \U \cr \cr \pmtheta^t \left({1 \over 2} \U^\prime +
\overline{\V}^\prime \right) + {1 \over 4} \U^t \U - \overline{\V}^t
\overline{\V}\endpmatrix  \tag{2.51}
$$
(which is the same as formula (2.42)).
Hence,
$$
D(\pmPhi_{(1)})^\dagger = \pmatrix \pmtheta && - {1 \over 2} \pmtheta
\partial + {1 \over 2} \U \cr \cr \o && - \pmtheta \partial - 2 \overline{\V}
\endpmatrix  \ .
\tag{2.52}
$$
\pvo \nd
Denoting, for brevity,
$$\tilde{u}: = \Phi_{(1)} (u) \ ,   \quad \tilde{h}: = \Phi_{(1)} (h) \ ,
\tag{2.53}
$$
we then get, for $H = uh/2$:
$$
\align
{\delta \Phi_{(1)} (H) \over \delta \U} &= \left ({1 \over 2}
\tilde{h} - {1 \over 4} \tilde{u}^\prime \right) \pmtheta + {1 \over 4}
\tilde{u} \U \ , \tag{2.54a}\\
{\delta \Phi_{(1)} (H) \over \delta \overline{\V}} &= - {1 \over 2}
\tilde{u}^\prime \pmtheta - \tilde{u} \overline{\V} \ .
\tag{2.54b}
\endalign
$$
Applying the Hamiltonian matrix $B_{(1)}$  (2.33) to the vector
(2.54a,b), we obtain the $\grg$-mDWW equations of motion:
$$
\align
\dot{\U} &= \left({1 \over 2} \tilde{h} - {1 \over 4}
\tilde{u}^\prime \right) [ \pmtheta, \U] + \left[ (\tilde{h} - {1 \over 2}
\tilde{u}^\prime) \pmtheta + {1 \over 2} \tilde{u} \U \right]^\prime
\tag{2.55a}\\
{\overline{\V}} &= {1 \over 2} \tilde{u}^\prime
[\overline{\V}, \pmtheta] + \left({1 \over 2} \tilde{u}^\prime \pmtheta +
{1 \over 2} \tilde{u} \overline{\V} \right)^\prime \ .
\tag{2.55b}
\endalign
$$

Similarly, in the Hamiltonian structure $\overline{B}_{(0)} \ (2.38)$,
the following sequence of Hamiltonians
$$
\H_1 = 2h, \quad \H_2 = {1 \over 2} (uh +h^2), \ldots \tag{2.56}
$$
is known to form an infinite commuting family {\bf [Ku1]}.  The first
flow, with Hamiltonian ${1 \over 2} \H_1 = h $, is
the $\sigma-$shift; the second flow, with  Hamiltonian $\H_2 = (uh +
h^2)/2$, gives
$$
\align
\dot{u}&= \left({1 \over 2} u^2 + uh - {1 \over 2}
u^\prime - h^\prime \right)^\prime  \tag{2.57a} \\
\noalign{\vskip 5 pt}
\dot{h}& = \left(uh + {3 \over 2} h^2 + {1 \over 2} h^\prime
\right)^\prime \ .  \tag{2.57b}
\endalign
$$
which are the first mDWW equations (in the variables $(u + 2h;h)$). (The
reader may check that in these variables one obtains the same system
as the $\{\grg = \Bbb R \}$-case of the system $(2.54a,b)$ in the variables
$(\U; \overline{\V} + {1 \over 2} \U)$).  The Hamiltonian map
$\Phi_{(0)}$ produces an infinite hierarchy of Hamiltonians
$\Phi_{(0)} (\H_n)$ commuting in the Hamiltonian structure $B_{(0)} \
(2.41)$. For the case $\grg = \Bbb R$, this hierarchy is known as the
(doubly modified) m$^2$DWW hierarchy.  Supppose now that $\grg$  is
arbitrary. The Hamiltonian
$$\
\Phi_{(0)} ({1 \over 2} \H_1) = \Phi_{(0)} (h ) \sim
\U^t \V  \tag{2.58}
$$
produces the $\sigma$-shift, as expected.  To compute the first nontrivial
flow, with Hamiltonian
$$
\Phi_{(0)} (\H_2) = \Phi_0 [{1 \over 2} (uh + h^2)] \ , \tag{2.59}
$$
we have:
$$
\pmPhi_{(0)} = \pmatrix \Phi_{(0)} (u) \cr \cr \Phi_{(0)} (h) \endpmatrix  =:
 \pmatrix \tilde{u} \cr \cr \tilde{h}\endpmatrix  = \pmatrix \pmtheta^t \U \cr
\cr
\pmtheta^t \V^\prime + \U^t \V \endpmatrix \ , \tag{2.60}
$$
(which is the same as (2.43a,b)) so that
$$
D(\pmPhi_{(0)})^\dagger = \pmatrix {\pmtheta}^t&&{\o}\cr\cr
\V^t&&\U^t + \pmtheta^t \partial\endpmatrix ^\dagger =
\pmatrix \pmtheta&&\V\cr\cr
\o&&\U- \pmtheta \partial\endpmatrix \ . \tag{2.61}
$$
Hence, for any $H \in C_{u, h}$,
$$\pmatrix \delta/\delta \U \cr \cr
\delta/\delta \V\endpmatrix \!\! \Phi_{(0)} (H) =
\pmatrix \pmtheta&&\V\cr \cr
\o&& \U- \pmtheta \partial
\endpmatrix
\!\! \Phi_{(0)}\!\! \pmatrix \delta H / \delta
u\cr\cr\delta H/ \delta h\endpmatrix  \ , \tag{2.62}
$$
and for the case $H = \H_2 = (uh + h^2)/2$, we get
$$
\pmatrix \dps{{\delta \Phi_0 (\H_2)} \strut \over \strut \dps{\delta
\U}}\cr \cr \cr
\dps{{\delta \Phi_0 (\H_2)} \strut \over \strut \dps{\delta \V}}\endpmatrix
= \pmatrix {1 \over 2} \tilde{h} \pmtheta + \left({1 \over 2} \tilde{u} +
\tilde{h}\right)\!\! \V \cr \cr \cr
\left({1 \over 2} \tilde{u} + \tilde{h}\right)\!\! \U -
\left({1 \over 2} \tilde{u} + \tilde{h}\right)^\prime\!\!
\pmtheta \endpmatrix  \ . \tag{2.63}
$$
Applying the Hamiltonian matrix $B_{(0)}$ (2.41) to the vector
(2.63), we get the first nontrivial flow in the $\grg$-m$^2$DWW hierarchy:
$$
\align \dot{\U}&= \left({1 \over 2}
\tilde{u} + \tilde{h}\right)^\prime [ \U, \pmtheta] +
\left[\left({1 \over 2} \tilde{u} +
\tilde{h}\right) \U - \left({1 \over 2} \tilde{u} +
\tilde{h}\right)^\prime\!\! \pmtheta \right]^\prime   \tag{2.64a}\\
\dot{\V}& = {1 \over 2} \tilde{h} [ \pmtheta,
\U] + \left({1 \over 2} \tilde{u} +
\tilde{h}\right)^\prime [\V, \pmtheta] + \left[{1 \over 2}
\tilde{h} \pmtheta + \left({1 \over 2} \tilde{u} +
\tilde{h}\right)\!\! \V \right]^\prime\ . \tag{2.64b}
\endalign
$$
{\it Remark 2.3}. \ There  also exists a family of {\it rational}
Hamiltonian maps (as opposed to the {\it polynomial} ones we have
been dealing with so far) from $(\grg \oplus \grg_{A})^{\wedge \ast}$
into $\text{diff}_{1 [ \gamma]}^{\ \wedge \ast}$,  $\Phi_{[\gamma]}: C_{u,h}
\rightarrow C_{\U, \V}$, of the form
$$
\align
\Phi_{[\gamma]} (u) &= \pmg^t \U - \epsilon
\sum_{s = 1}^N \left[ \text{ln}(\U + \V)^t \pmtheta^s \right]^\prime
\tag{2.65a}\\
\Phi_{[\gamma]} (h) &= {1 \over 4} ( \U^t \U - \V^t\V) \ , \tag{2.65b}
\endalign
$$
where $\pmtheta^1, \ldots, \pmtheta^N$ is a family of pairwise commuting
constant elements in $\grg$, $\pmg$ is another constant element in $\grg,
\ \gamma: = \pmg^t \pmg$, and
$$
\align
B &= \pmatrix 2 \partial \b1&&\o\cr\cr
\o&& -4[\V, \ ]-2 \partial \b1\endpmatrix  \tag{2.66a}\\
\overline{B}_{[\gamma]} &=
\pmatrix 2 \pmg^t \pmg \partial&&\partial
u - \partial^2\cr\cr
u \partial + \partial^2&&h \partial + \partial h\endpmatrix \ .
\tag{2.66b}
\endalign
$$
In particular, when $\pmg^t \pmg = 1$,
the Hamiltonian matrix $\overline{B}_{[1]}$  of eq\. (2.66b) is the same as
$\overline{B}_{(1)}$in (2.36).  Therefore, the
set of Hamiltonians $\{\Phi_{[1]} (H_n)\}$
forms a new $\grg$-mDWW infinite commuting hierarchy.  Similarily, when
$\pmg^t \pmg = 0$,
the Hamiltonian matrix $\overline{B}_{[0]}$  of  (2.66b) is the same as
the Hamiltonian matrix $\overline{B}_{(0)}$ in  (2.38).  Hence, the set
of Hamiltonians $\{\Phi_{[0]} (\H_n)\}$
forms a new $\grg$-m$^2$DWW infinite commuting hierarchy.  The geometric
nature of the map $\Phi_{[\gamma]}$  defined by eqs\. (2.65a,b) is as yet
 a mystery.
\bigskip
\noindent
\line{2c. \quad {\it  Specializations}\hfill}
\medskip
The hierarchy of DWW equations
$$
\pmatrix u \cr \cr h \endpmatrix _{\!\! t} = \overline{B}_{(1)}
\pmatrix \delta H_m/\delta u \cr \cr \delta H_m/\delta h\endpmatrix  \ , \ \ \
m \in \N \ , \tag{2.67}
$$
has, for every  odd flow $m = 1 (\text{mod } 2)$,
the invariant submanifold defined by
$u = 0$, on which this hierarchy reduces to the KdV hierarchy
$$
 h_t = {1 \over 2} \partial \left({\delta \overline{H}_{2m+3}
\over \delta h} \right) = ({1 \over 2} \partial^3 + h \partial +
\partial h) \left({\delta \overline{H}_{2 m + 1} \over \delta h}
\right) \ , \tag{2.68}
$$
where
$$
\overline{H}_m : = H_m \bigm |_{u = 0} \ . \tag{2.69}
$$
This follows from the following formulae {\bf[Ku1]}:
$$
\align
{\delta \overline{H}_{2m} \over \delta h} &= 0  \tag{2.70a}
\\
{\delta \overline{H}_{m +1} \over \delta h} &= 2 {\delta H_m \over
\delta u} \biggm|_{u = 0} - \partial \left({\delta \overline{H}_m \over
\delta h} \right)
\tag{2.70b}\\
\partial \left({\delta H_{m +1} \over \delta u}\right) \Biggl |_{u =
0}& = (h \partial + \partial h) \left({\delta \overline{H}_m \over
\delta h}\right) + \partial^2 \left({\delta H_m \over \delta u}\right)
\Biggr|_{u = 0}
\tag{2.70c}\\
\partial \left({\delta \overline{H}_{m +2} \over \delta h}\right) &=
[2(h \partial + \partial h) + \partial^3] \left({\delta \overline{H}_m
\over \delta h}\right)
\tag{2.70d}\\
\partial \left({\delta H_{m + 2} \over \delta u} \right) \Biggl |_{u
= 0} &= [ 2 (h \partial + \partial h) + \partial^3] \left({\delta H_m
\over \delta u} \right) \Biggr|_{u = 0} + (h^\prime \partial + \partial
h^\prime) \left({\delta \overline{H}_m \over \delta h} \right)
\tag{2.70e}\\
\pmatrix u \cr \cr h\endpmatrix _{\!\! t}& = \overline{B}_{(1)} \pmatrix \delta
H_m/
\delta u \cr \cr \delta H_m/ \delta h\endpmatrix  = \pmatrix 0&&\partial\cr \cr
\partial&&0\endpmatrix  \pmatrix \delta H_{m +1} / \delta u \cr \cr \delta
H_{m + 1} /\delta h\endpmatrix  \ .
\tag{2.71}
\endalign
$$
Thus,  when
$$m = 2n + 1, \ \ \ n \in \Z_+\ , \tag{2.72}
$$
from formulae (2.71), (2.70a), we get
$$
 u_t \Bigl |_{u = 0} = \partial ({\delta H_{2n +2} \over
\delta h}) \Bigr|_{u = 0} \ = 0 \ ,  \tag{2.73}
$$
so that $\{u = 0 \}$ is indeed an invariant submanifold.  From
formulae (2.71) and (2.70b,c), we obtain
$$
\align
 h_t \Biggl|_{u = 0} &=  \partial \left({\delta H_{2n + 2} \over
\delta u}\right) \Biggr|_{u = 0}
= \partial {1 \over 2} \left({\delta \overline{H}_{2 n + 3} \over
\delta h} \right) \\
&=\left(h \partial + \partial h +
{1 \over 2} \partial^3 \right)\!\! \left({\delta \overline{H}_{2n + 1} \over
\delta h} \right)  \ ,
\endalign
$$
which is eq\. (2.68).

Similar results can be derived for  the nonabelian integrable systems of
the preceding subsection.
\proclaim
{Theorem 2.3}\ {\it (i)} \  The $\grg$-mDWW system
$$
\pmatrix \U \cr \cr \overline{\V}\endpmatrix _{\!\! t} = \pmatrix -2 [ \U, \ \
]
+
\partial \b1&\o\cr \cr \o&-[\overline{\V}, \ ] - {\dps{1} \strut \over
\strut \dps{2}}
\partial \b1\endpmatrix  \pmatrix \delta/\delta \U \cr \cr \delta/\delta
\overline{\V} \endpmatrix \!\! \Phi_{(1)} (H_{2n + 1})
\tag {2.74}
$$
has the invariant submanifold defined by
$$
\U = \o   \ ,\tag{2.75}
$$
on which it becomes
$$
\overline{\V}_{,t} = \left( - [ \overline{\V}, \ \ ] - { 1 \over 2} \partial
\b1 \right)\!\! {\delta \over \delta \overline{\V}} \left( \Phi_{(1)}
(\overline{H}_{2n + 1}) \right) \ . \tag{2.76}
$$
{\it (ii)} \ The map $\Phi_{(1)}$ on the submanifold $\{ \U = \o \}$:
$$
\overline{\Phi}_{(1)} (h) : = \Phi_{(1)} (h) \bigm|_{\U = \o} =
\pmtheta^t \overline{\V}^\prime - \overline{\V}^t \overline{\V} \tag{2.77}
$$
is a Hamiltonian map into the second Hamiltonian structure of the
$KdV$ hierarchy (2.68):
$$
h_t = (h \partial + \partial h + {1 \over 2} \partial^3)\!\!
\left({\delta \overline{H}_{2 n + 1} \over \delta h} \right)\ .
\tag{2.78}
$$
\endproclaim
\demo {Proof}
{\it (i)}\quad  By formulae $(2.50)$ and $(2.46)$,
$$
\align
{ \delta \Phi_{(1)} (H_{2 n +1}) \over \delta \U} \Biggl|_{\U =
\o} &= \pmtheta^t \Phi_{(1)}\!\! \left({ \delta H_{2n +1} \over \delta u} - { 1
\over 2} \partial \left({\delta H_{2n + 1} \over \delta h} \right)
\right) \Biggr|_{u = 0}  \\
\text{[by (2.70b)]}\qquad &= \pmtheta^t \Phi_{(1)}\!\! \left( {1 \over 2}
{\delta \overline{H}_{2n +2}
\over \delta h} \right) \\
 \text{[by (2.70a)]}\qquad & = \pmtheta^t \Phi_{(1)}(0) = \o \ .
\tag{2.79}
\endalign
$$
Hence, from (2.74),
$$
\2 \U_t \Biggl|_{\U = \o} = \partial \left({\delta \Phi_{(1)} (H_{2
n + 1}) \over \delta \U} \Biggr|_{\U = \o} \right) = \o \ , \tag{2.80}
$$
so that $ \U = \o $ indeed defines an invariant submanifold.

\noindent
{\it (ii)} \quad  From formulae $(2.50)$ and $(2.52)$ we have
$$
\2{\delta \Phi_{(1)} ( \overline{H}_{2n +1} ) \over \delta
\overline{\V}} = {\delta \Phi_1 (H_{2n + 1}) \over \delta
\overline{\V}} \biggm|_{\U = \o } = - (\pmtheta \partial + 2 \overline{\V})
\Phi_{(1)}\!\! \left({\delta
\overline{H}_{2n + 1} \over \delta h} \right)\ , \tag{2.81}
$$
since letting $\U$ vanish does not interfere with taking variational
derivatives with respect to $\overline{\V}$.  Hence, from the second
row of formula $(2.74)$ we conclude that
$$
\overline{\V}_t \biggm|_{\U = \o} = \left( - [ \overline{\V},
\ ] - {1 \over 2} \partial \b1 \right)\!\! \left({\delta \Phi_{(1)} (
\overline{H}_{2n+1}) \over \delta \overline{\V}} \right) \ ,
$$
which is $(2.76)$.  Then a computation similar to the one at the end
of subsection 2a shows that
$$
\overline{\Phi}_{(1)} \left(h \partial + \partial h + {1 \over 2}
\partial^3 \right) = D \left[ \overline{\Phi}_{(1)} (h) \right]
\left(- [ \overline{\V}, \ ]
- {1 \over 2} \partial \b1 \right) \{D[\overline{\Phi}_{(1)} (h)]\}^\dagger
\ .
\tag{2.82}
$$
\vskip-16pt
\hfill $\square$
\enddemo
 It follows from Theorem 2.2 that  condition (2.75) picks out the flows
of eq\. (2.76), which are  the $\grg$-mKdV flows constructed in
{\bf [Ku2]}.

The problem of specialization is slightly different for the
$\grg$-m$^2$DWW hierarchy, defined by
$$
\pmatrix \U \cr \cr \V \endpmatrix _{\!\! t} = \pmatrix \o&&-[\U, \ \ ] +
\partial \b1 \cr \cr -[\U, \ \ ] + \partial \b1&&-[\V, \ \ ]\endpmatrix
\pmatrix \delta/\delta \U \cr \cr \delta/ \delta \V \endpmatrix \!\! \Phi_{(0)}
(\H_m)\ .   \tag{2.83}
$$
To begin with, the hierarchy related to it by the map $\Phi_{(0)}$ :
$$
\2 \pmatrix u \cr \cr h\endpmatrix _{\!\! t} = \pmatrix 0&& \partial u -
\partial^2 \cr \cr u \partial + \partial^2&&h \partial + \partial h
\endpmatrix  \pmatrix \delta/\delta u \cr \cr \delta/\delta h\endpmatrix
 (\H_m)    \tag{2.84}
$$
has an invariant submanifold defined by
$$
u + 2h = 0  \tag{2.85}
$$
for all $m \equiv 1 (\text{mod}\ 2)$.  This  follows from the equality
$$
\hskip 1.00truein {\delta \H_{2n +1} \over \delta h} = 2
{\delta \H_{2n +1} \over \delta u} \qquad {\text{on \ }}  \{u + 2h = 0
\}  \ .  \tag{2.86}
$$
(see {\bf [R]}).  Fixing $n$ and defining
$$
\cX_n: = {\delta \H_{2n +1} \over \delta u} \biggm|_{u + 2h = 0} \  ,
 \tag{2.87}
$$
we obtain from formula (2.62) that on the submanifold defined by
$$
\Gamma: = \pmtheta^t(\U + 2 \V^\prime) + 2 \U^t \V = 0 \ , \tag{2.88}
$$
which is the image under the homomorphism $\Phi_{(0)}$  (2.43a,b) of the
invariant submanifold  determined by $u + 2h = 0 $, we have
$$
\align
 {\delta \Phi_{(0)} (\H_{2n + 1}) \over \delta \U}
\biggm|_{\Gamma = 0} &= ~\cX_n (\pmtheta + 2 \V) \tag{2.89a}\\
 {\delta \Phi_{(0)} (\H_{2n + 1}) \over \delta \V}
\biggm|_{\Gamma = 0} &= 2 \cX_n \U - 2 \cX_n^\prime \pmtheta \ . \tag{2.89b}
\endalign
$$
Hence, on the submanifold defined by  $\Gamma = 0 $, our
 $\grg$-m$^2$DWW system (2.83) becomes
$$
\align
\U_t &= 2 \cX_n^\prime [ \U, \pmtheta] + (2 \cX_n \U - 2
\cX_n^\prime \pmtheta)^\prime  \tag{2.90a}\\
\V_t &= \cX_n [ \pmtheta, \U] + 2 \cX_n^\prime [ \V, \pmtheta] +
[\cX_n(\pmtheta + 2 \V)]^\prime \ .  \tag{2.90b}
\endalign
$$
\proclaim
{Proposition 2.4}  The flow  determined by (2.90) leaves the
submanifold  defined by $\Gamma = 0 $ invariant.
\endproclaim
\demo{Proof}  From (2.90a,b), we obtain
$$
\Gamma_t = 2 \cX_n \Gamma^\prime + 4 \cX_n^\prime \Gamma \ .
\tag{2.91}
$$
\vskip-16pt \hfill $\square$
\enddemo
\noindent
{\it Remark 2.3}.  The constraint $\Gamma = 0 $  in (2.88)
is differential and cannot be resolved algebraically, except in the
classical case when $\grg = \Bbb {R}^1$ and $\pmtheta = 1$, resulting in the
formula
$$
U = - (1 + 2V)^{-1} V^\prime = \left[- {1 \over 2} \text{ln} (1 + 2V)
\right]^\prime \ , \tag{2.92}
$$
so that one is dealing with the potential mKdV hierarchy.

\bigskip
\line{2d. \quad{\it Integrable Systems on $T^*\wt G$}\hfil}
\medskip
{}From formula (1.10d) for the vector field  $X_H$ of the Hamiltonian
$H(g, \mu)$, the equations of motion may be expressed as
$$
\pmatrix
\dot g \\ \\
\dot \mu
\endpmatrix
= B^k
\pmatrix
\frac {\delta H }{ \delta g} \\ \\
 \frac {\delta H }{ \delta \mu}
\endpmatrix
\ ,
\tag{2.93}
$$
where the Hamiltonian matrix $B^k$ is
$$
B^k=
\pmatrix
0 &&  L_{g*}  \\  \\
- R_g^{*} && [\mu + k g^{-1}g',  \ . \ ] + k \partial
\endpmatrix \ , \tag{2.94}
$$
and the notation $L_{g*}$ and $R_g^{*}$ signifies left (resp\. right)
multiplication  of the $\grg$ element
$\frac {\delta H }{\delta \mu}$ (resp\. the  $T^*_g\wt{G}$ element
$\frac {\delta H }{ \delta g}$) by $g$. The integrable commuting hierarchies
constructed in the previous subsections generate commuting hierarchies
on $T^*\wt{G}$ by pulling back those on
$\text{diff}_{1(\a)}^{\ \wedge *}$ under the Poisson maps (1.76, 1.108)
of Section 1.

For the case $(\a)=(0)$, the commuting system of Hamiltonians is
generated by $\{({\hat J}^{LA})^*\phi_{(0)}(\Cal H_n)\}$.
For $n=1$, by formula (2.58),
we have $\Phi_{(0)} ({1 \over 2} \Cal H_1) = \Phi_{(0)} (h ) \sim \U^t \V$.
Since, by formula (1.46),
$$
\align
({\hat J}^{LA})^{*}(\U) &= g'g^{-1}  \tag{2.95a} \\
({\hat J}^{LA})^{*}(\V) &= g \mu g^{-1}  \ ,  \tag{2.95b}
\endalign
$$
we find that
$$
\align
\frac {\d}{\d g} (\hat J^{LA})^*(\U^t \V)  &=
 (-\mu' + [\mu, g^{-1}g'] )g^{-1} \tag{2.96a} \\
\frac{\d}{\d \mu}(\hat J^{LA})^*(\U^t \V)  &=
g^{-1}g'  \ , \tag{2.96b}
\endalign
$$
and the equations of motion (2.93), (2.94) yield
$$
\pmatrix \dot g \\ \\
\dot{\mu}
\endpmatrix =
\pmatrix g' \\ \\
\mu'
\endpmatrix \ ,
\tag{2.97}
$$
which is just the $\sigma$-shift, as expected.
For all the other Hamiltonians $\Cal H_n$, we let:
$$\
\d \Phi_{(0)} (\Cal H_n)=: \wt{a}^t_n \d \U + \wt{b}^t_n \d \V \ .  \tag{2.98}
$$
Thus, e.g., $\wt{a}_2$ and $\wt{b}_2$ are given by the components of the
vector (2.63).
Denoting
$$
a_n:= (\hat J^{LA})^*(\wt a_n),\quad b_n:=(\hat J^{LA})^*(\wt b_n) \ ,
\tag{2.99}
$$
we find that
$$
\align
\frac {\d}{\d g} (\hat J^{LA})^* \Phi_{(0)}(\Cal H_n) &=
g^{-1}[g\mu g^{-1}, b_n] - (g^{-1}a_n)'- g^{-1}a_ng'g^{-1}  \tag{2.100a} \\
\frac {\d}{\d \mu} (\hat J^{LA})^* \Phi_{(0)}(\Cal H_n) &=
 g^{-1}b_n g \ , \tag{2.100b}
\endalign
$$
so that the equations of motion on $T^*\wt G$ for the Hamiltonian
$(\hat J^{LA})^* \Phi_{(0)}(\Cal H_n)$ are
$$
\align
\dot g &= b_n g  \tag{2.101a}\\
\dot \mu & = -(g^{-1}a_n g)' \ .   \tag{2.101b}
\endalign
$$

Next, consider the case $(\alpha)=(1)$. Here we get the infinite commuting
hierarchy $\{(\hat J^{LR}_k)^* \Phi_{(1)}(H_n)\}$, where $\hat J^{LR}_k$ is
given by formula (1.23):
$$
\align
(\hat J^{LR})^{*}(\overline\U) &= g\mu g^{-1} \ , \quad  \overline \U :=
{1\over 2} \U
 \tag{2.102a} \\
(\hat J^{LR})^{*}(\overline\V) &= -\mu + k g^{-1}g' \ ,
\quad k= {1 \over 2} \ .
 \tag{2.102b}
\endalign
$$
For the Hamiltonian $H_1 =h$, we have from (2.42b) that
$\Phi_{(1)} (H_1) \sim \overline \U^t \overline\U - \overline{\V}^t
\overline{\V}$, so that
$$
\align
\d \Phi_{(1)} (H_n) & = :\wt{a}^t_n \d \overline \U +
 \wt{b}^t_n \d \overline \V \ ,  \tag{2.103}\\
\wt{a}_1 = 2 \overline \U, & \quad \wt b_1 = -2 \overline \V \ .  \tag{2.104}
\endalign
$$
Denoting
$$
\wh{a}_n := (\hat J^{LR})^{*}(\wt a_n),
\quad \wh{b}_n := (\hat J^{LR})^{*}(\wt b_n) \ , \tag{2.105}
$$
from formulae (2.102a,b) we get
$$
\align
\frac {\d}{\d g} (\hat J^{LR}_k)^* \Phi_{(1)}( H_n) &=
g^{-1}[g\mu g^{-1},  \wh a_n] - k(\wh b_n' +[g^{-1}g', \wh b_n])g^{-1} \ ,
  \tag{2.106a} \\
\frac {\d}{\d \mu} (\hat J^{LR}_k)^* \Phi_{(1)}(H_n) &=
 -\wh b_n + g^{-1}\wh a_n g \ , \tag{2.106b}
\endalign
$$
and formula (2.93) yields the following Hamiltonian equations on
$T^{*}\wt G$ for the Hamiltonian  $(\hat J^{LR}_k)^* \Phi_{(1)}(H_n)$:
$$
\align
\dot g &= \wh a_n g - g \wh b_n \ ,  \tag{2.107a}\\
\dot \mu & = k (g^{-1}\wh a_n g)' + [\wh b_n, \mu] + k [g^{-1}g', g^{-1}\wh a_n
g],
\quad k= {1 \over 2} \ . \tag{2.107b}
\endalign
$$
In particular for $n=1$, substituting formulae (2.104) and (2.102a,b) in
(2.107a,b), we get the $\sigma$-shift flow, as expected.
\bigskip

\line{2e. \quad{\it WZW Model} \ {\bf [Wi,  FT, H]}\hfil}
\medskip
For the $3$-parameter family of Hamiltonians on $T^*\wt G$
$$
H= \int \left[\a_1 \langle \mu, \mu \rangle + \a_2 \langle \a_2,
g^{-1}g'\rangle
+ \a_3 \langle g^{-1}g', g^{-1}g' \rangle  \right]  d\s \ ,  \tag{2.108}
$$
we have
$$
\align
\frac{\d H}{\d g} &= - \lbrace (2\a_3 g^{-1}g' + \a_2 \mu)' +
 [g^{-1}g', \a_2 \mu] \rbrace g^{-1}  \tag{2.108a} \\
\frac{\d H}{\d \mu} &= 2 \a_1 \mu + \a_2 g^{-1}g' \ .   \tag{2.108b}
\endalign
$$
The corresponding equations of motion, by formula (2.93), are therefore
$$
\align
\dot g &= 2 \a_1 g \mu + \a_2 g'  \tag{2.109a} \\
\dot \mu &= \lbrace (2\a_3 + \a_2 k)g^{-1}g' + (\a_2 + 2 \a_1 k)\mu \rbrace'
  + 2\a_1 k [g^{-1}g', \mu]  \ .  \tag{2.109b}
\endalign
$$
Eliminating  the momentum $\mu$ from these equations gives the
second order system
$$
{1 \over 2\a_1} (g^{-1}\dot{g})\dot{}
= (2\a_3 + {k \over 2} \a_2)(g^{-1}g')' +
 {k \over 2} [g^{-1}\dot{g},g^{-1}g'] \ .  \tag{2.111}
$$
Hence, choosing
$$
\a_1 = {\eps \over k}, \quad \a_2 = - \eps, \quad \a_3 = {\eps k \over 2},
\quad \eps = \pm 1 \ , \tag{2.112}
$$
we obtain the standard WZW system {\bf [Wi]}
$$
 (g^{-1}\dot{g})\dot{} - (g^{-1}g')' =
\eps[g^{-1}\dot{g},g^{-1}g'] \ .  \tag{2.113}
$$

The Hamiltonian $H$ in (2.108) for these choices of the parameters
 $\a_1$, $\a_2$, $\a_3$ is:
$$
H_{WZW} = {\eps \over 2k} \int \left[ \langle\mu, \mu \rangle +
\langle - \mu + k g^{-1}g', - \mu + k g^{-1}g'\rangle \right] d\s \ ,
 \tag{2.114}
$$
which  by formulae (2.102a,b), with $k= {1 \over 2}$ and $\eps =1$, is
$$
(\wh J^d)^{*} \left( \int \left[{\U^t\U \over 4} +
\overline{\V}^t \overline{\V}\right]d\s  \right)  \ , \tag{2.115}
$$
i.e., the pull back of the Hamiltonian
$$
\wh H_{WZW} = \int \left[{\U^t\U \over 4} +
\overline{\V}^t \overline{\V}) \right] d\s
= \int  \left[(\overline{\U}^t \overline{\U}  +
\overline{\V}^t \overline{\V}\right]d\s
\tag{2.116}
$$
from $(\wt \grg \oplus \wt \grg)^{\wedge *}$. The corresponding equations
of motion on $(\wt \grg \oplus \wt \grg)^{\wedge *}$ are, by formula
(2.33), just the usual left and right translational modes
$$
\align
\dot{\overline{\U}} &=\overline{\U}'   \tag{2.117a} \\
\dot{\overline{\V}} &= -\overline{\V}' \ . \tag{2.117b}
\endalign
$$
Since the system (2.117a,b) has integrals of the form
$$
F_1(\overline{\U}) + F_2(\overline{\V}) \ ,  \tag{2.118}
$$
with arbitrary $\s$-independent Hamiltonians $F_1$ and $F_2$, the WZW
system (2.109) has an infinite number of integrals of the form
$$
\int \left[ F_1(g\mu g^{-1}) +
 F_2(-\mu + {1 \over 2} g^{-1}g') \right] d\s \ . \tag{2.119}
$$
These integrals commute, for distinct pairs  $(F_1, F_2)$ and
$(\tilde{F}_1, \tilde{F}_2)$, if $(F_1, \tilde{F}_1)$ and
$(F_2, \tilde{F}_2)$ separately do. Thus, we may choose them from
any two commuting hierarchies on $\wt{\grg}^{\wedge*}$. There is a
distinguished such hierarchy; namely, the  nonabelian mKdV
hierarchy {\bf[Ku2]},  obtained by pulling back the KdV integrals
under the map $\Phi$ of (2.27) with an arbitrary fixed $\pmtheta$ of
length $1$:
$$
\Phi(h_n) =: h_n^{\pmtheta}  \ . \tag{2.120}
$$
Hence, we have the doubly infinite commuting system
on  $(\wt\grg \oplus \wt \grg)^{\wedge *}$ consisting of Hamiltonians
of the form $\{ h_n^{\pmtheta_1}(\overline{\U}) +
 h_m^{\pmtheta_2}(\overline{\V})\}$.
 This, in turn, furnishes the KdV-generated
doubly infinite commuting hierarchy of Hamiltonians on $T^{*}\wt{G}$ of
the form $\{ h_n^{\pmtheta_1}(g\mu g^{-1}) +
 h_m^{\pmtheta_2}(-\mu + {1 \over 2} g^{-1}g')\}$,
 which include the generator of  the left-right translational flow giving the
WZW model as its first element.
\bigskip

\newpage
\centerline{\smc References}
\bigskip
{\smaller{
\item{\bf [AHP]} Adams, M.R., Harnad, J. and Previato, E., ``Isospectral
Hamiltonian Flows in Finite and Infinite Dimensions I. Generalised Moser
Systems
and Moment Maps into Loop Algebras'',  {\it Commun\. Math\. Phys\.} {\bf 117},
451-500 (1988); Adams, M.R., Harnad, J. and Hurtubise, J., ``Isospectral
Hamiltonian Flows in Finite and Infinite Dimensions II.  Integration of
Flows'',
 {\it Commun\. Math\. Phys\.} {\bf 134}, 555-585 (1990).
\item{\bf [DJKM]}
Date, E\., Kashiwara, M\., Jimbo, M. and Miwa, T\.,
``Transformation Groups for Soliton Equations'',
I\. {\it Proc\. Jap\. Acad\.} {\bf 57A}, 342-347 (1981);
II\. {\it ibid\.}, 387-392 (1981); III\. {\it J\. Phys\. Soc Japan}, {\bf 50},
3806-3812 (1981); VI\. {\it ibid\.} {\bf 50} 3813-3818 (1981);
V\. {\it Publ. Res\. Inst\. Math\. Sci\.} {\bf 18}, 1077-1110 (1982);
{\it ibid\.}, 1111-1119 (1982); IV\. {\it Physica} {\bf 4D}, 343-365 (1982).
\item{\bf [DS]}
Drinfeld, V\.~G\. and Sokolov, V\.~V\.,
``Lie algebras and Equations of Korteweg - De Vries Type'',
{\it Jour. Sov. Math.} {\bf 30} 1975-2036 (1985);
``Equations of Korteweg-De Vries type and Simple Lie Algebras'',
{\it Soviet Math\. Dokl\.} {\bf 23}, 457-462 (1981).
\item{\bf [FNR]}
Flaschka, H\., Newll, A\.C\. and Ratiu, T\.,
``Kac-Moody Algebras and Soliton Equations II\. Lax Equations Associated
to $A_1^{(1)}$'',
{\it Physica} {\bf 9D} 300 (1983).
\item{\bf [FT]} Faddeev, L.D\. and Takhtajan, L\.A\.,
  {\it Hamiltonian Methods in the Theory of Solitons}, Part II, Ch.1.5,
 Springer-Verlag, Heidelberg (1987).
\item{\bf [H]}
Harnad, J\.,
``Constrained Hamiltonian systems on Lie Groups, Moment Map Reductions
and Central Extensions'', preprint CRM (1990).
\item{\bf [HK1]}
Harnad, J\. and Kupershmidt, B\. A\.,
``Twisted Diff $S^1$-action on Loop Groups and
Representations of the Virasoro Algebra'',
{\it Lett\.  Math\. Phys\.} {\bf 19}, 277-284 (1990).
\item{\bf [HK2]}
Harnad, J\. and Kupershmidt, B\. A\.,
``Hamiltonian Group Actions on Superloop Spaces'',
{\it Commun\.  Math\. Phys\.} {\bf 132}, 315-347 (1990).
\item{\bf [Ku1]}
 Kupershmidt, B\. A\.,
``Mathematics of Dispersive Water Waves'',
{\it Commun\.  Math\. Phys\.} {\bf 99}, 51-73 (1985).
\item{\bf [Ku2]}
 Kupershmidt, B\. A\.,
``Modified Korteweg-de Vries Equations on Euclidean Lie Algebras'',
{\it Int\. J\. Mod\. Phys\.} {\bf 3}, 853-861 (1989).
\item{\bf [PS]} Pressley, A\. and Segal, G\.,
{\it Loop Groups}, Clarendon Press, Oxford (1986).
\item{\bf [RS]} Reiman, A.~G. and Semenov-Tian-Shansky, M.~A.,
``Reductions of Hamiltonian Systems, Affine Lie  Algebras and Lax
 Equations I and II'',
{\it Invent\. Math\.} {\bf 54}, 81-100 (1979); {\it ibid\.} {\bf 63},
423-432 (1981).
\item{\bf [R]}
 Razboinick, S\. I\.,
``Vector Extensions of Modified Water Wave Equations'',
{\it Phys\. Lett\.} {\bf 119A}, 283-286 (1986).
\item{\bf [SW]}
 Segal, G\. and Wilson, G\.,
``Loop Groups and Equations of KdV Type'',
{\it Publ\. Math\. IHES} {\bf 61}, 5-65 (1985).
\item{\bf [W1]}
 Wilson, G\.,
``Habillage et fonctions $\tau$'',
{\it C\. R\. Acad\. Sci\. Paris, S\'er. I} {\bf 299}, 587-590 (1984).
\item{\bf [W2]}
 Wilson, G\.,
``On the Quasi-Hamiltonian Formalism of the KdV Equation'',
{\it Phys\. Lett\.} {\bf 132A}, 445-450 (1988).
\item{\bf [Wi]}
Witten, E\.,
``Nonabelian Bosonization in Two Dimensions'',
{\it Commun\. Math\. Phys\.} {\bf 92}, 452-472 (1984).
\item{}
}}
%
\enddocument